\documentclass[traditabstract]{aa}

\usepackage[varg]{txfonts}
\usepackage{bm}
\usepackage{graphicx}
\usepackage{latexsym}
\usepackage{color}
\usepackage{amssymb}
\usepackage{subfigure}
\usepackage{supertabular}
\usepackage{natbib,twoopt}
\usepackage{ulem}

\usepackage[breaklinks=true]{hyperref} 
\bibpunct{(}{)}{;}{a}{}{,} 
\makeatletter
\newcommandtwoopt{\citeads}[3][][]{\href{http://adsabs.harvard.edu/abs/#3}%
{\def\hyper@linkstart##1##2{}%
\let\hyper@linkend\@empty\citealp[#1][#2]{#3}}}
\newcommandtwoopt{\citepads}[3][][]{\href{http://adsabs.harvard.edu/abs/#3}%
{\def\hyper@linkstart##1##2{}%
\let\hyper@linkend\@empty\citep[#1][#2]{#3}}}
\newcommandtwoopt{\citetads}[3][][]{\href{http://adsabs.harvard.edu/abs/#3}%
{\def\hyper@linkstart##1##2{}%
\let\hyper@linkend\@empty\citet[#1][#2]{#3}}}
\newcommandtwoopt{\citeyearads}[3][][]%
{\href{http://adsabs.harvard.edu/abs/#3}
{\def\hyper@linkstart##1##2{}%
\let\hyper@linkend\@empty\citeyear[#1][#2]{#3}}}
\makeatother

\def\kelvin{\,\mathrm{K}}

\begin{document}

\title{Scaling of collision strengths
    for highly-excited states of ions of the H- and He-like sequences
       \thanks{Tables of atomic data for \ion{Si}{xiii} and \ion{S}{xv}
         are only available in electronic form at the CDS via anonymous
         ftp to \url{ftp://cdsarc.u-strasbg.fr} (130.79.128.5)
         or via \url{http://cdsarc.u-strasbg.fr/viz-bin/qcat?J/A+A/}
      }
   }

\author{L. Fern\'{a}ndez-Menchero\inst{1,2} \and
     G. Del~Zanna\inst{3} \and
     N.~R. Badnell\inst{1}
    }
\institute{Department of Physics, University of Strathclyde. 
        Glasgow G4 0NG, United Kingdom 
        \email{luis.fernandez-menchero@drake.edu} \and
        Present address: Department of Physics and Astronomy, Drake University. 
        2507 University Avenue. Des Moines, IA 50311, United States \and
        Department of Applied Mathematics and Theoretical Physics, 
        University of Cambridge, Cambridge CB3 0WA, 
        United Kingdom
       }

\abstract{Emission lines from highly-excited states ($n \ge 5$) of $\mathrm{H}$- and
$\mathrm{He}$-like ions have been detected in astrophysical sources and
fusion plasmas. 
For such excited states,  $R$-matrix or distorted wave calculations for 
electron-impact excitation are very limited, 
due to the large size of the atomic basis set needed to describe them.
Calculations for $n \ge 6$ are also not generally available.
We study the behaviour of the  electron-impact excitation collision strengths and effective 
collision strengths for the most important transitions used to model electron 
collision dominated astrophysical plasmas, solar, for example.
We investigate the dependence on the relevant parameters: the principal quantum 
number $n$ or the nuclear charge $Z$.
We also estimate the importance of coupling to highly-excited states and the continuum by comparing
the results of different sized calculations.
We provide analytic formulae to calculate the electron-impact excitation 
collision strengths and effective collision strengths to highly-excited states 
($n \ge 8$) of $\mathrm{H}$- and $\mathrm{He}$-like ions.
These extrapolated effective collision strengths can be used to interpret
astrophysical and fusion plasma via  collisional-radiative modelling.
}
\authorrunning{L. F. Menchero et al.}
\titlerunning{Scaling of collision strengths for highly-excited atomic states}

\maketitle

\section{Introduction}
\label{sec:introduction}

Spectral emission lines  of H- and He-like ions have been used for diagnostics 
of both fusion and astrophysical plasmas for decades.
Perhaps the most famous examples are the temperature and 
density diagnostics of the He-like ions in the X-rays, described by 
\cite{jordan1969}.
However, the status of the atomic data for these ions still requires 
improvement.
As described below, atomic data for ions in these sequences are generally
only available up to the principal quantum number $n=5$. 
Atomic data for  highly-excited levels are needed for a variety of reasons.
First, lines from these levels (up to $n=10$) have been observed in 
laboratory plasma (see, e.g. the compilations in the 
NIST database \citealt{nist2013})
and recently also in  X-ray spectra of solar flares
(see, e.g. \citealt{kepa2006}).
Second, transitions between highly-excited levels should be included for any 
appropriate  collisional-radiative  modelling of these ions.
Third, even if in most astrophysical spectra  lines from these levels
are not readily visible, they do contribute  to the X-ray 
pseudo-continuum, so they should be included in any spectral modelling.

Calculating atomic data for highly-excited levels is not a trivial task and 
has various limitations, since it requires a significant increase in the
size of the atomic basis set.
In Section~\ref{sec:atomic} we review previous calculations for the electron-impact 
excitation of He- and H-like ions, and present the results of test calculations 
made with larger basis sets.
These calculations are performed to see how well the extrapolated data 
agree with the calculated ones for higher $n$.
In Section~\ref{sec:extrapolation} we study  the behaviour of the  electron-impact 
excitation collision strengths and effective collision strengths for 
several kinds of transitions.
These are the most common transitions that decay producing the lines 
observed in astrophysics.
We also estimate the importance of coupling to highly-excited states and the continuum
by comparing the results of differently sized distorted wave and $R$-matrix calculations.
We then provide analytic formulae to calculate electron-impact excitation 
collision strengths to highly-excited states ($n \ge 8$) of $\mathrm{H}$- 
and $\mathrm{He}$-like ions.
This is done by extrapolating the results obtained with the $R$-matrix or 
distorted wave methods.
Potentially, the method provides results  up to $n=\infty$, although accuracy reduces as $n$ increases.
In Section~\ref{sec:comparison} we compare the solar flare  line intensities 
with those predicted by applying the extrapolation rules to the effective collision
strengths. 
Finally, in Section~\ref{sec:conclusions} we summarise the main conclusions.

\section{Atomic data}
\label{sec:atomic}

A number of calculations for the electron-impact 
excitation of ions of the $\mathrm{H}$- and $\mathrm{He}$-like sequences
can be found in the literature.
Authors have used a number of different methodologies and different 
configuration interaction (CI) and close coupling (CC) basis sets.

\citet{whiteford2001} calculated electron-impact excitation effective
collision strengths for  $\mathrm{He}$-like ions.
\citet{whiteford2001} included in the CI / CC basis set all the 
single-electron excitations up to principal quantum number $n=5$ 
(49 fine-structure levels)
and used the radiation-damped intermediate coupling frame transformation ICFT $R$-matrix 
method \citep{griffin1998}.
These data are the most rigorous and complete to date.
They can be found in the UK APAP network 
database\footnote{www.apap-network.org}, as well as in 
OPEN-ADAS\footnote{open.adas.ac.uk} and in the most recent 
version 8 \citep{delzanna2015c} of the CHIANTI 
database\footnote{www.chiantidatabase.org}.

There are other studies in the literature.
\citet{aggarwal2005} calculated electron-impact effective 
collision strengths for $\mathrm{Ar}^{16+}$ up to $n=5$ using the Dirac 
$R$-matrix method~\citep{norrington1987}.
\citet{chen2006} calculated Dirac $R$-matrix electron-impact
excitation effective collision strengths for $\mathrm{Ne}^{8+}$ up 
to $n=5$.
\citet{kimura2000} performed Dirac $R$-matrix calculations for the
$\mathrm{He}$-like ions $\mathrm{S}^{14+}$, $\mathrm{Ca}^{18+}$ 
and $\mathrm{Fe}^{24+}$ up to $n=4$.

In the $\mathrm{H}$-like sequence, \citet{ballance2003} performed a detailed
study of hydrogenic ions from $\mathrm{He}^{+}$ to $\mathrm{Ne}^{9+}$ 
plus $\mathrm{Fe}^{25+}$. 
\citet{ballance2003} used a quite extensive basis set, including pseudostates.
The basis set included all the spectroscopic terms up to
$n=5$ for all the ions except $\mathrm{Ne}^{9+}$.
For $\mathrm{Ne}^{9+}$ the basis set was extended up to $n=6$.
The pseudo-state terms included in the calculations varied  for each ion.

Even though the above calculations are quite extensive, they are 
still insufficient for the modelling of highly-excited shells ($n > 5$),
as noted in the introduction.

In the present work we have performed some test calculations with an 
extensive basis set up to $n=8$.
The calculations are focused on checking the validity of the extrapolation methods
that we have developed, and discuss in the next section. As such, we do not
include radiation damping of resonances.
The target basis set includes all the possible $l$ values for $n=1-6$, and then
up to $\mathrm{7g}$ and $\mathrm{8f}$.
We have performed both $R$-matrix and distorted wave  calculations with
the same basis set.
The $R$-matrix suite of codes are described in \cite{hummer1993} 
and \cite{berrington1995}.
The calculation in the inner region was in $LS$ 
coupling and included mass 
and Darwin relativistic energy corrections.
The outer region calculation used the ICFT method \citep{griffin1998}.
The distorted wave calculations were carried-out using the
{\sc autostructure} program \citep{badnell2011b}.
The ICFT $R$-matrix and distorted wave calculations were carried out with the same atomic
structure to estimate the effects of the resonances and coupling in general.

To estimate the collision strengths for higher shells 
($n=8-12$) we performed a different
distorted wave calculation.
In this second calculation we used a configuration basis set 
consisting of $\mathrm{1s}^2$  and $\mathrm{1s}nl$ for $n=8-12$ and $l$ up to
$\mathrm{8h}$, $\mathrm{9h}$, $\mathrm{10g}$, $\mathrm{11f}$ and
$\mathrm{12d}$, that is, we neglect configurations with $n=2-7$.
Thus, although we automatically have CI between these more highly-excited $n$-shells, 
there is no mixing with lower ones, save for the ground.
The atomic structure is oversimplified in order to get a description of the highly-excited 
states which becomes increasingly demanding when retaining the full CI expansion.
This calculation has a poorer atomic structure so it is expected that these results will 
not be of such high accuracy.
It has been performed only to check if such an oversimplified atomic structure
can give results for the effective collision strengths with an error which is
acceptable for plasma modelling, 
and to compare that error with the one arising from the extrapolation of results obtained
using the (smaller) full-CI expansion.

\section{Extrapolation rules}
\label{sec:extrapolation}

The scattering calculations provide the collision strengths $\Omega$ 
as a function of the incident electron energy.
The collision strengths are extended to high energies by interpolation using  
the appropriate high-energy limits in the  \cite{burgess1992} scaled domain.
The infinite energy limit points are calculated with {\sc autostructure}.
The temperature-dependent effective collisions strength $\Upsilon$ are 
calculated by convoluting these collision strengths with a Maxwellian electron velocity distribution.

The behaviour of the collision strengths $\Omega$ and effective
collision strengths $\Upsilon$ for highly-excited levels follows the
semi-empirical formula:
\begin{equation}
\Omega(n),\Upsilon(n)\ \sim\ \frac{A}{(n+\alpha)^3} \,,
\label{eq:extrap}
\end{equation}
where $n$ is the principal quantum number and $A$ and $\alpha$ are parameters 
to be determined.
Usually $\alpha$ is small and can be set to zero.
The formula \ref{eq:extrap} is usually a good description for highly-excited states,
where the atom can be considered as a Rydberg one, meaning that for $n$ at least two
units more than the last atomic shell occupied by any inactive core electrons
of the ion.

We used three models to determine the parameters of the formula \ref{eq:extrap}:
\begin{description}
\item[Model 1:] Least-squares fit, including all the calculated values
    of $\Omega$ and $n$.
\item[Model 2:] Two point extrapolation, calculate $A$ and $\alpha$
    from the values of $\Omega$ and $n$ of the last two points calculated.
\item[Model 3:] One point extrapolation, fix $\alpha=0$ and calculate $A$
    from the value of $\Omega$ and $n$ of the last point calculated.
\end{description}
We have compared the results of these extrapolation Models with those
obtained from explicit $R$-matrix and distorted 
wave test calculations which we have performed.
In the following sections we discuss the  different cases.

\subsection{Dipole-allowed transitions}
\label{subsec:dipole}

These transitions are between states with opposite parity and with changes 
in total angular momenta $\Delta J=0, 1$ (and $J=0\rightarrow 0$ forbidden)
that is to say, electric dipole.
The collision strength diverges logarithmically as the collision energy tends to infinity.
\citep{burgess1992}.
In general, the contribution from resonances will be small compared to the background, so
distorted wave and $R$-matrix methods produce similar results for the effective collision
strengths.

\begin{figure*}[!hbtp]
\centering
\subfigure{
  \includegraphics[width=0.45\textwidth]{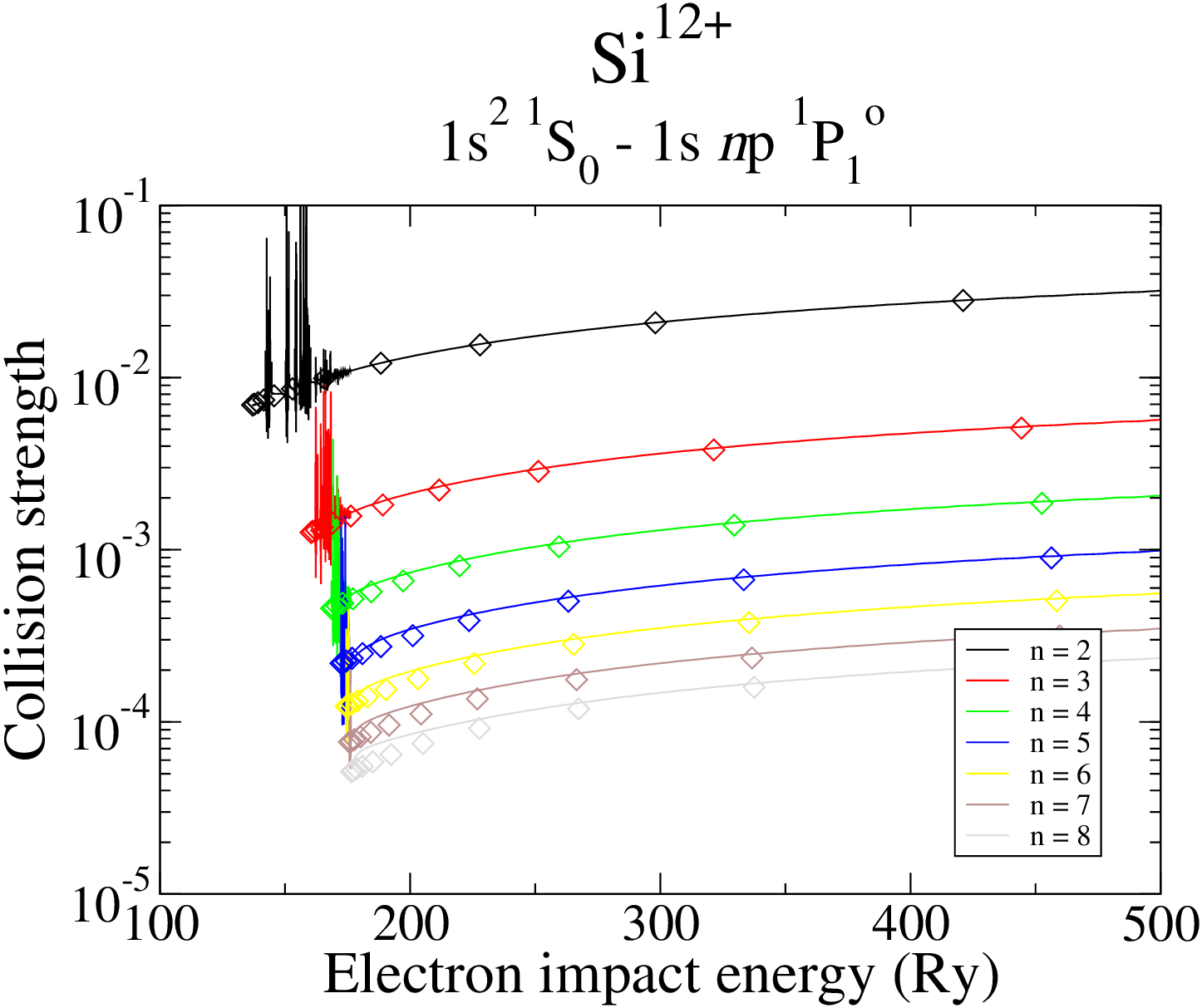}
}\,
\subfigure{
  \includegraphics[width=0.45\textwidth]{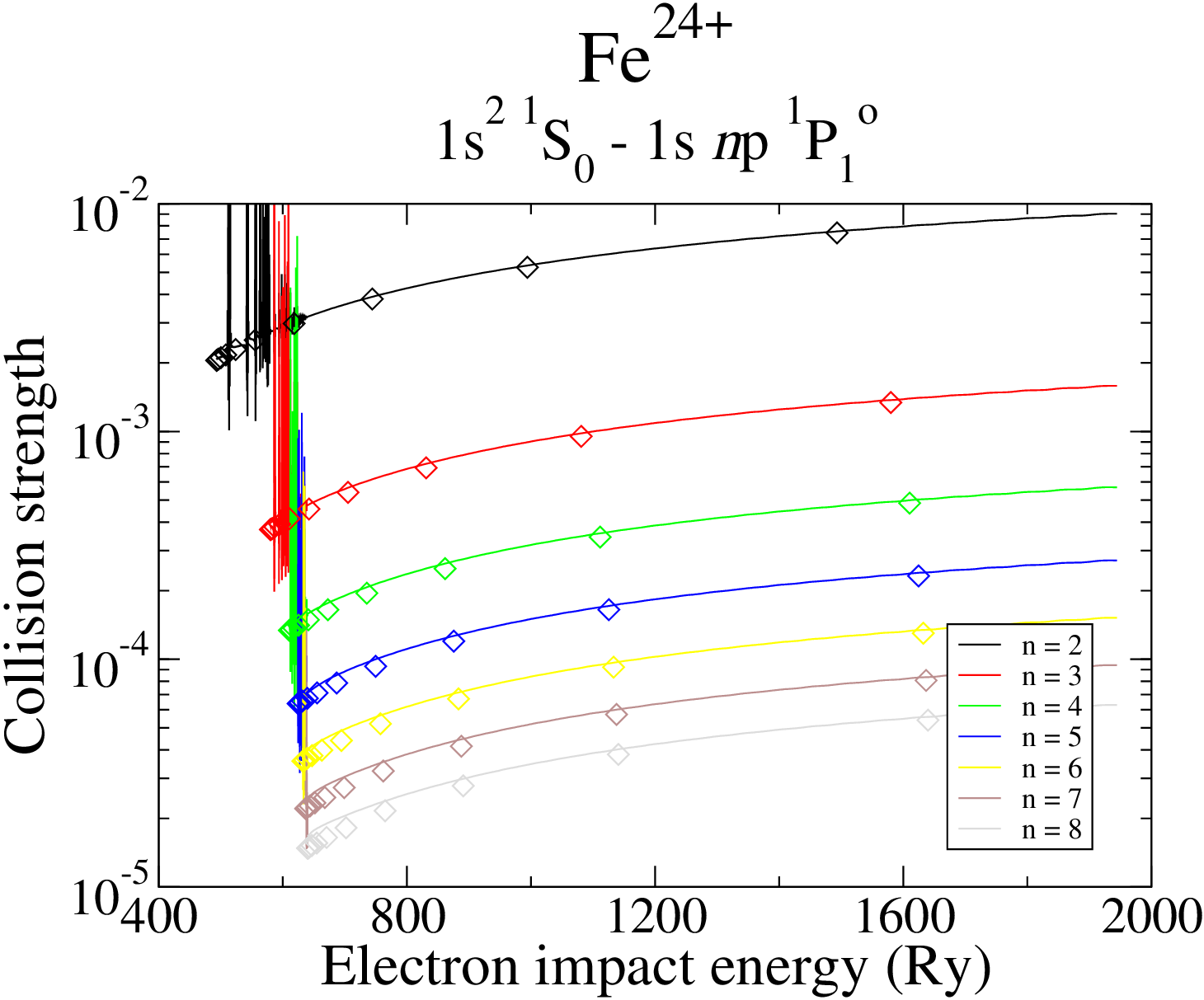}
}\\
\caption{Electron-impact excitation collision strengths for 
  the electric dipole transition 
  $\mathrm{1s^2\,^1S_0}-\mathrm{1s}n\mathrm{p\,^1P_1}$
  of the ions $\mathrm{Si}^{12+}$ and $\mathrm{Fe}^{24+}$.
  Curved lines: $R$-matrix; $\diamond$: Distorted Wave calculation.}
\label{fig:omg_np1P}
\end{figure*}

Figure~\ref{fig:omg_np1P} shows the calculated collision strengths for the 
electric dipole transitions 
$\mathrm{1s^2\,^1S_0}-\mathrm{1s}n\mathrm{p\,^1P_1}$ of the
moderately-charged  $\mathrm{Si}^{12+}$ and the highly-charged 
$\mathrm{Fe}^{24+}$ ions.
We plot both the results of the  $R$-matrix and distorted wave methods.
Both calculations  were carried out with the same atomic structure, with a somewhat large 
CI/CC expansion, up to the atomic shell $n=8$.

\begin{figure*}[hbtp]
\centering
\subfigure{
  \includegraphics[width=0.45\textwidth]{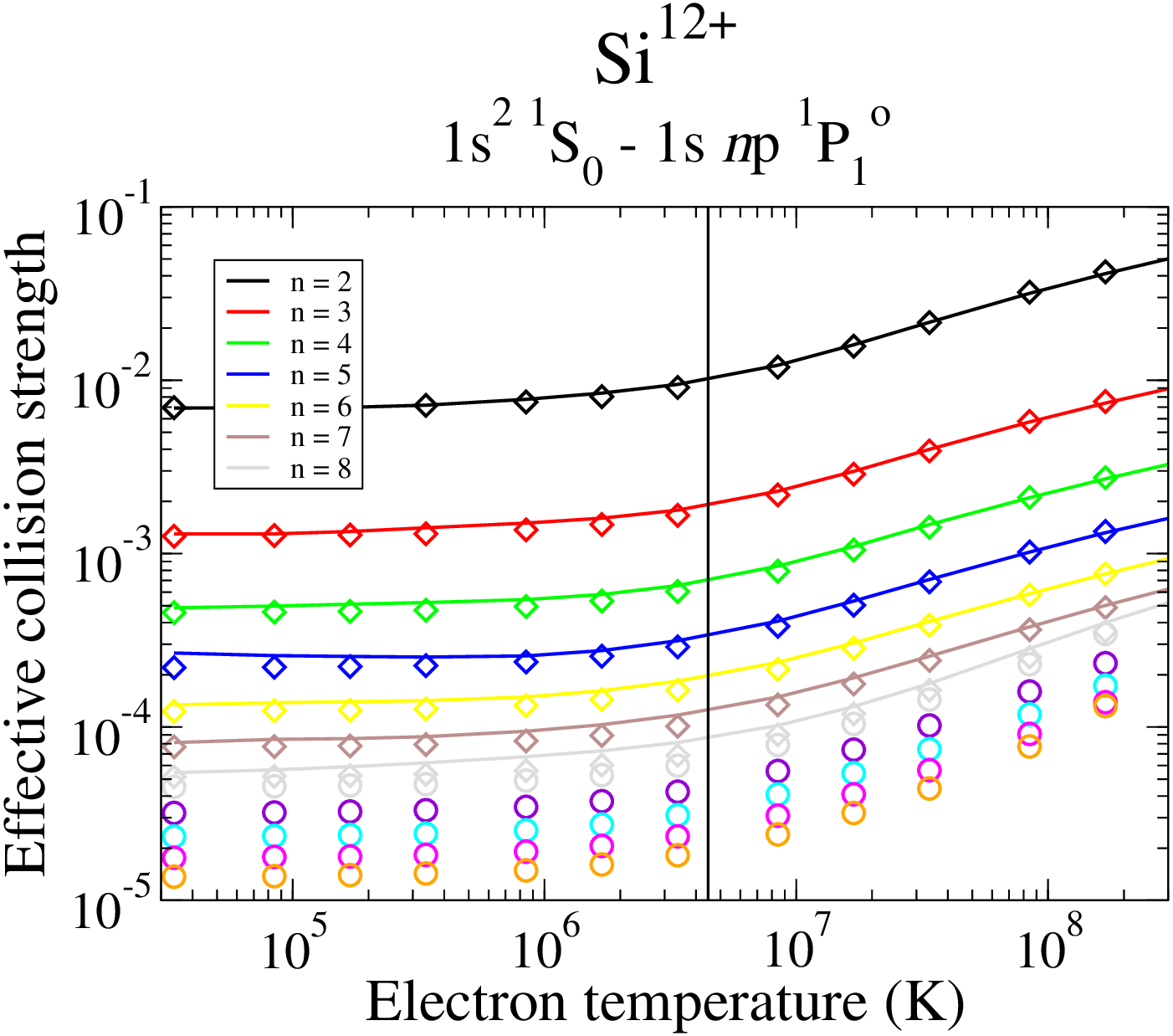}
}\,
\subfigure{
  \includegraphics[width=0.45\textwidth]{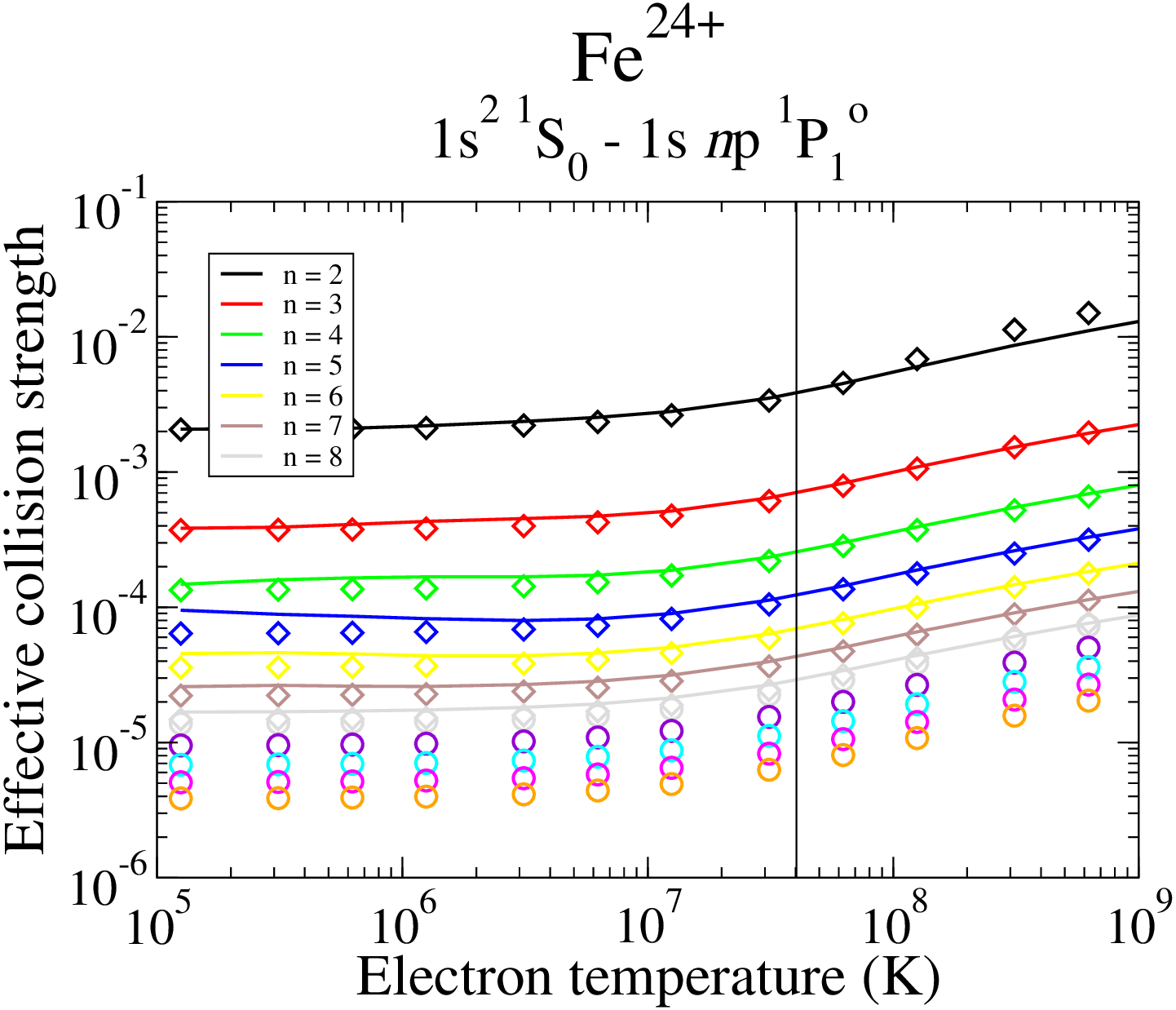}
}\\
\caption{Electron-impact excitation effective collision strengths for 
  the electric dipole transition 
  $\mathrm{1s^2\,^1S_0}-\mathrm{1s}n\mathrm{p\,^1P_1}$
  of the ions $\mathrm{Si}^{12+}$ and $\mathrm{Fe}^{24+}$.
  Curved lines: $R$-matrix; 
  $\diamond$: distorted wave calculation basis $n=1-8$;
  $\circ$: distorted wave calculation basis $n=8-12$;
  vertical line: peak abundance temperature for electron collisional 
  plasmas~\citep{mazzotta1998}.}
\label{fig:ups_np1P}
\end{figure*}

At low temperatures, the effects of the resonances is not significant.
This is expected for strong dipole transitions, where the background is large
in comparison.
As we pointed out in \cite{fernandez-menchero2015b}, the $R$-matrix 
calculations can not guarantee accuracy at very low temperatures, of the order
of the energy of the first excited level.
In fact, uncertainties associated with the position of the resonances can 
reach $100\%$ for such low temperatures.
However, for electron collision dominated plasmas, for example solar, the ions are 
mainly formed near the peak abundance temperature (vertical lines in the plots
in Fig.~\ref{fig:ups_np1P}),
and at these temperatures the position of the
resonances has a negligible effect on the effective collision strength.

Figure~\ref{fig:ups_np1P} shows the Maxwell-integrated effective collision 
strengths for the same transitions and ions.
The difference in  the $\Upsilon$ between both calculations for the transition 
from the ground level to $\mathrm{8p}$ is around $15\%$.

\begin{figure*}[hbtp]
\centering
\subfigure{
  \includegraphics[width=0.45\textwidth]{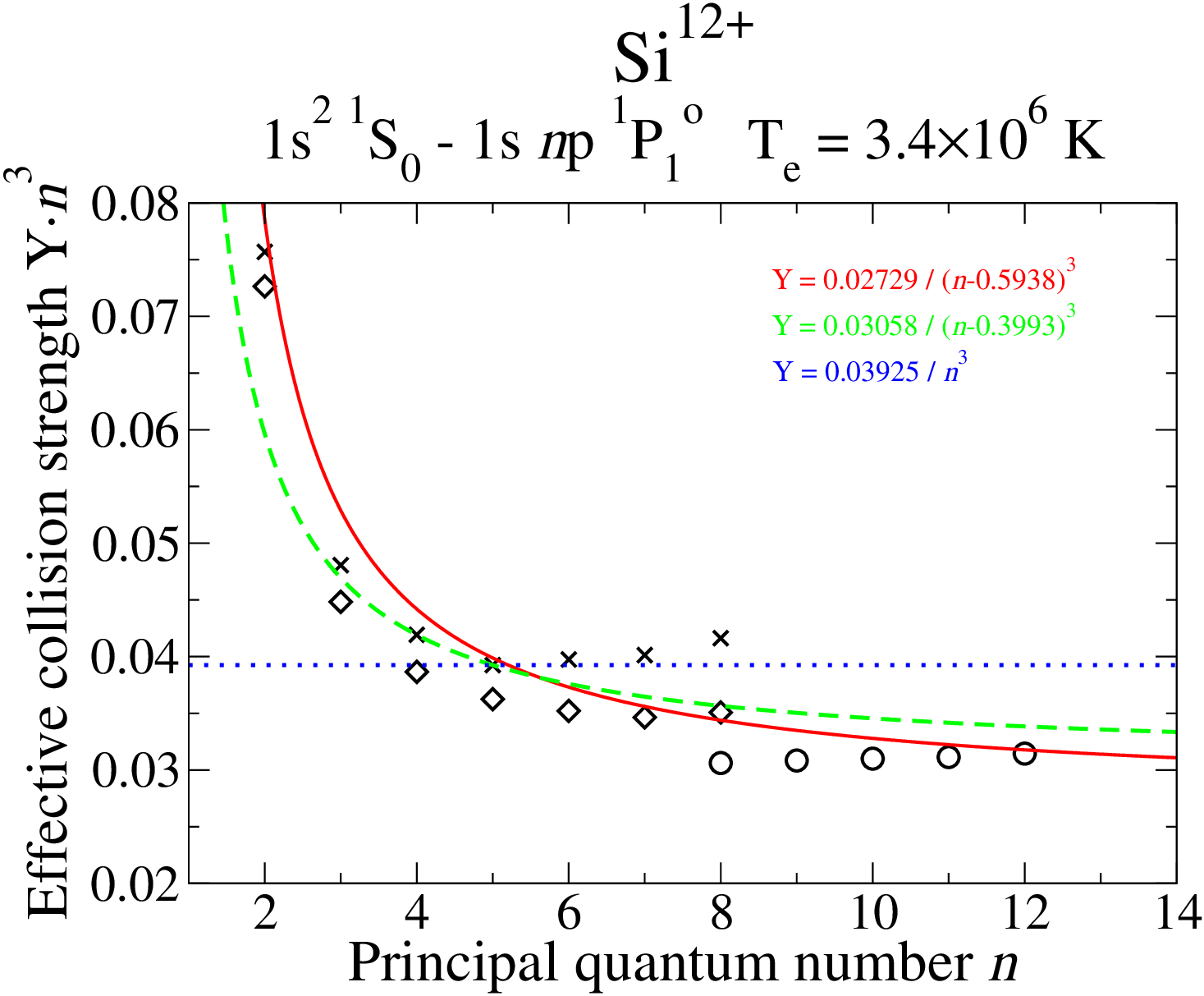}
}\,
\subfigure{
  \includegraphics[width=0.45\textwidth]{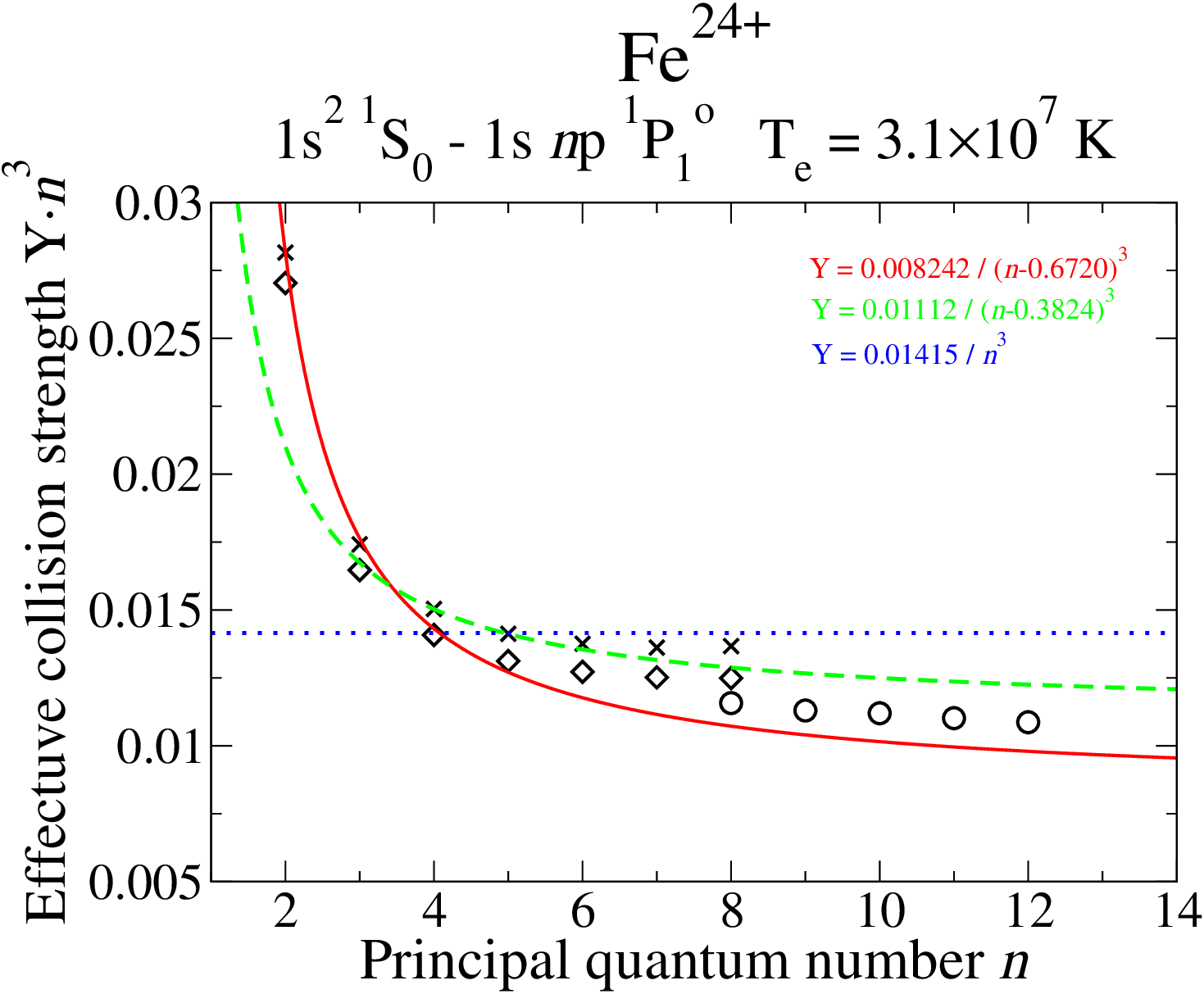}
}\\
\caption{Electron-impact excitation effective collision strengths  $(\times\, n^3)$ for 
  the electric dipole transition 
  $\mathrm{1s^2\,^1S_0}-\mathrm{1s}n\mathrm{p\,^1P_1}$
  of the ions $\mathrm{Si}^{12+}$ and $\mathrm{Fe}^{24+}$ 
  versus the principal quantum number $n$ around peak abundance temperature.
  $\times$: $R$-matrix results;
  $\diamond$: distorted wave results with basis set $n=1-8$;
  $\circ$: distorted wave results with basis set $n=8-12$;
  solid line: least-squares fit using points $n=2-5$;
  dashed line: extrapolation using the last two points;
  dotted line: extrapolation using the last point; see text.}
\label{fig:upsvsn_np1P}
\end{figure*}

Figure \ref{fig:upsvsn_np1P} shows the behaviour of the effective collision
strengths $\Upsilon$ with respect to the principal quantum number $n$ at the
peak abundance temperature.
We compare the extrapolations from $n=5$ with the calculated values for the
three models.
For the lower-charged ion, $\mathrm{Si}^{12+}$, the disagreement between the
$R$-matrix and distorted wave results increases more rapidly for higher $n$,
reaching $20\%$ at $n=8$.
This is due to stronger coupling between the more highly-excited states
included in the close-coupling expansion.
However, we note that the $R$-matrix calculation cannot accurately describe transitions
to the highest states included in the CI/CC expansion \citep{fernandez-menchero2015b}.
For $\mathrm{Fe}^{24+}$, the $R$-matrix and distorted wave results agree better 
with each other, to $10\%$ at $n=8$, as coupling decreases with increasing charge.

\begin{figure}[hbtp]
\centering
\includegraphics[width=0.45\textwidth]{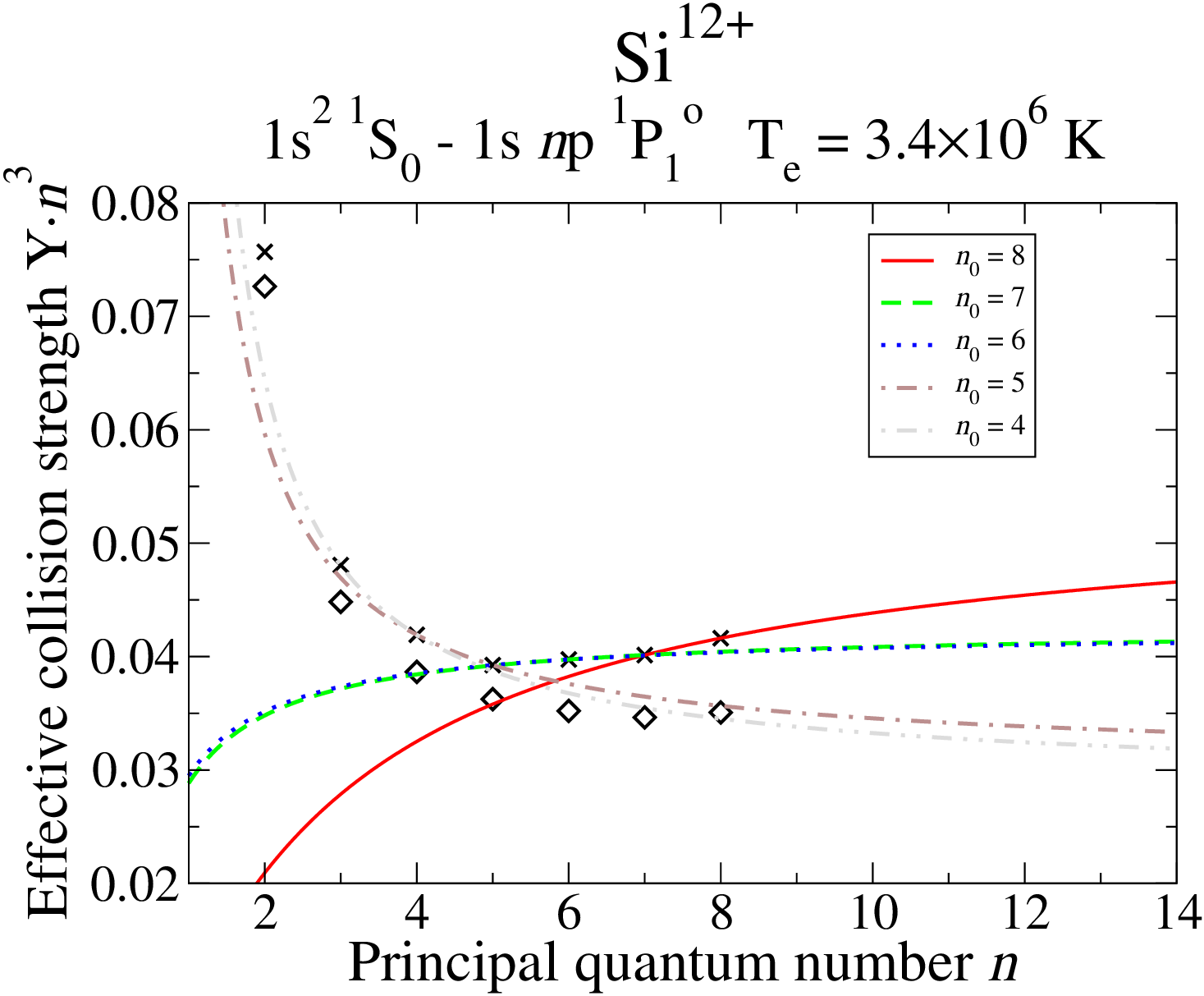}
\caption{Extrapolation curves for the $\Upsilon\times\, n^3$ displayed in 
  fig. \ref{fig:upsvsn_np1P}, taking different extrapolation
  points $n_0$.
  $\times$: $R$-matrix results;
  $\diamond$: distorted wave results.}
\label{fig:upsvsnmulti_np1P}
\end{figure}

Table \ref{tab:params_np1P} shows the different extrapolation parameters
calculated with the three methods for $\mathrm{Si}^{12+}$, and choosing different
reference points $n_0$ for the extrapolation.
The linear fit is performed taking into account all the points between $n=2$ 
and $n_0$.
The two-point model takes the values of $\Upsilon(n)$ for $n=n_0-1$ and $n_0$
and calculates the parameters $A$ and $\alpha$ through a two-equation
and two-unknown system (\ref{eq:extrap}).
Finally, the one-point model uses $\Upsilon(n_0)$ to calculate $A$ and 
sets $\alpha=0$.
Fig. \ref{fig:upsvsnmulti_np1P} displays the calculated analytic 
functions corresponding to each of the three models.
The predicted extrapolation curves with the two-point model vary considerably
in terms of the reference point $n_0$, by more than $30\%$.
They are influenced too much by the smaller values of $n$, which have yet
to reach their asymptotic form.
We see a similar variation in the two-point model.
On the other hand, the one-point extrapolation gives more stable results.
The predicted value of $A$ changes by just $10\%$ with the different choices of $n_0$.

With the linear fit it is necessary to include many points to obtain acceptable statistics and to reduce the associated error of kind $\beta$. 
In a linear least-squares fit, an acceptable number of points is around twelve
to get a $\beta$ error under $20\%$.
The computation cost of the linear fit is also larger.
For such a small set of points, the linear fit is not an appropriate model.
Thus, for strong dipole electric transitions we recommend use of the one-point extrapolation.

The last calculated $n$-shell $(n=8)$ is not a good reference point for the extrapolation
due to the lack of convergence of the CI and CC expansions, compared to $n\le 7$.
The parameters $A$ and $\alpha$ calculated with the two-point model are very 
similar using the second and the third last point as reference, and they are
also with $\alpha$ closest to zero.
These two curves for $n_0=6$ and $n_0=7$ shown in Fig. \ref{fig:upsvsnmulti_np1P} 
are the best extrapolation models for this type of transition, if such data is available.
If not, for smaller values of $n_0$, the one-point extrapolation is the best model.

\begin{table}[hbt]
\caption{\label{tab:params_np1P} Fitting parameters for the extrapolation
of the $\Upsilon$ at high-$n$ for the dipole electric transition 
of $\mathrm{Si}^{12+}$ $\mathrm{1s^2\,^1S_0}-\mathrm{1s}n\mathrm{p\,^1P_1}$
at a temperature of $T=3.4 \times 10^6 \kelvin$,
for different extrapolation reference points $n_0$.}
\begin{center}
\begin{footnotesize}
\begin{tabular}{|r|rr|rr|r|}
\hline
$n_0$ & \multicolumn{2}{c|}{Linear fit} &
\multicolumn{2}{c|}{Two point} & \multicolumn{1}{c|}{One point} \\
  & \multicolumn{1}{c}{$A$} & \multicolumn{1}{c|}{$\alpha$} &
    \multicolumn{1}{c}{$A$} & \multicolumn{1}{c|}{$\alpha$} &
    \multicolumn{1}{c|}{$A$} \\
\hline
  4 & $0.02560$ & $-0.5938$ & $0.02877$ & $-0.4715$ & $0.04192$  \\ 
  5 & $0.02729$ & $-0.5491$ & $0.03058$ & $-0.3993$ & $0.03925$  \\ 
  6 & $0.03004$ & $-0.4693$ & $0.04234$ & $ 0.1279$ & $0.03974$  \\ 
  7 & $0.03241$ & $-0.3959$ & $0.04257$ & $ 0.1387$ & $0.04013$  \\ 
  8 & $0.03509$ & $-0.3080$ & $0.05449$ & $ 0.7513$ & $0.04163$  \\ 
\hline
\end{tabular}
\end{footnotesize}
\end{center}
\end{table}

\subsection{Born-allowed transitions}
\label{susec:born}

For non-dipole Born-allowed transitions the collision strength tends to a constant value as
the collision energy tends to infinity, 
given by the plane-wave Born $\Omega_{\infty}^{\mathrm{PWB}}$ \citep{burgess1992}.
The $\Omega_{\infty}^{\mathrm{PWB}}$ is zero for double-electron jumps and 
spin-change transitions, in the absence of mixing.
In the intermediate coupling scheme (IC), most  transitions will be
Born-allowed or dipole-allowed through configuration and/or spin-orbit mixing.

\begin{figure*}[!hbtp]
\centering
\subfigure{
  \includegraphics[width=0.45\textwidth]{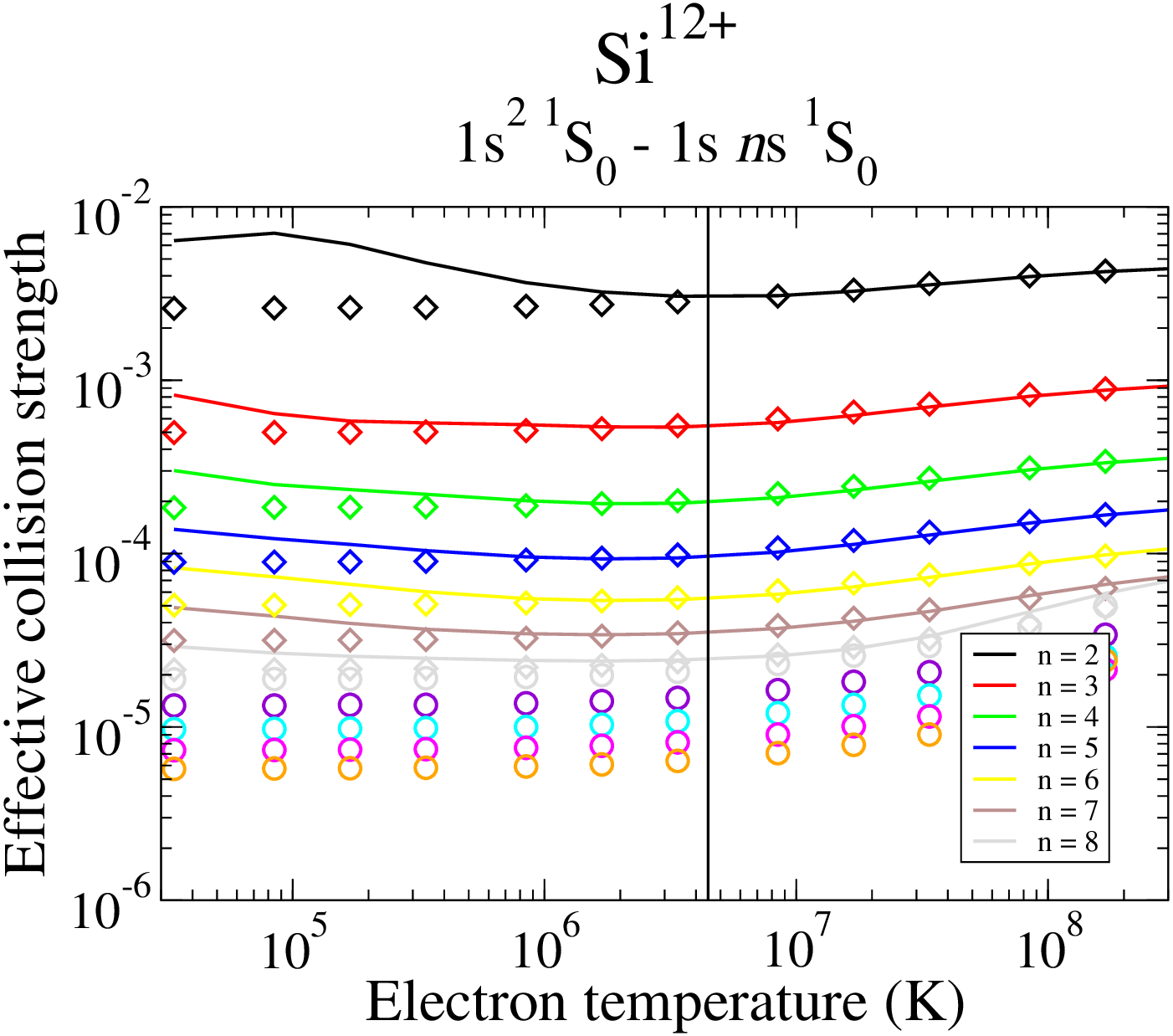}
}\,
\subfigure{
  \includegraphics[width=0.45\textwidth]{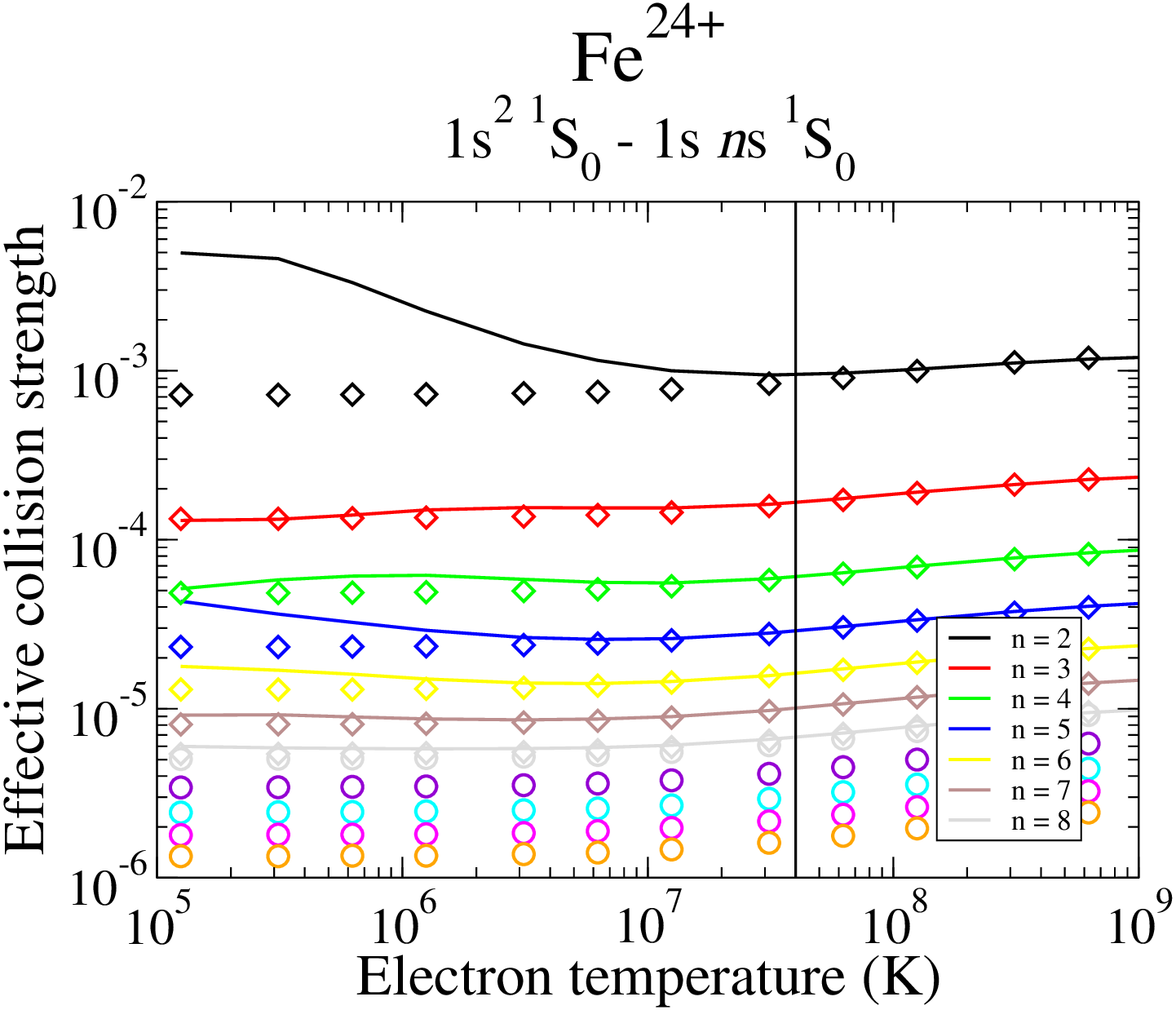}
}\\
\caption{Electron-impact excitation effective collision strengths for 
  the Born transition 
  $\mathrm{1s^2\,^1S_0}-\mathrm{1s}n\mathrm{s\,^1S_0}$
  of the ions $\mathrm{Si}^{12+}$ and $\mathrm{Fe}^{24+}$.
  Curved line: $R$-matrix; 
  $\diamond$: distorted wave calculation basis $n=1-8$;
  $\circ$: distorted wave calculation basis $n=8-12$;
  vertical line: peak abundance temperature for electron collisional 
  plasmas~\citep{mazzotta1998}.}
\label{fig:ups_ns1S}
\end{figure*}

Figure~\ref{fig:ups_ns1S} shows the effective collision strengths for the
one-photon optically forbidden transitions 
$\mathrm{1s^2\,^1S_0}-\mathrm{1s}n\mathrm{s\,^1S_0}$ for the ions
$\mathrm{Si}^{12+}$ and $\mathrm{Fe}^{24+}$.
This kind of transition has a very weak background collision strength at all energies,
so the enhancement due to resonances is large.
This effect is largest at low temperatures; the $\Upsilon$ calculated using
the distorted wave method are a considerable underestimate 
compared to those obtained with the $R$-matrix method.
The underestimation for the lowest temperatures ($\sim 10^{5} \kelvin$) and lowest
excited states can reach a factor of between two and ten.
This effect is reduced progressively at higher temperatures and for more highly
excited states.
At the peak abundance temperature, the resonance enhancement is reasonably small,
reduced to $10\%$ for a moderately-charged or a highly-charged ion, see figure~\ref{fig:ups_ns1S}.

\begin{figure*}[hbtp]
\centering
\subfigure{
  \includegraphics[width=0.45\textwidth]{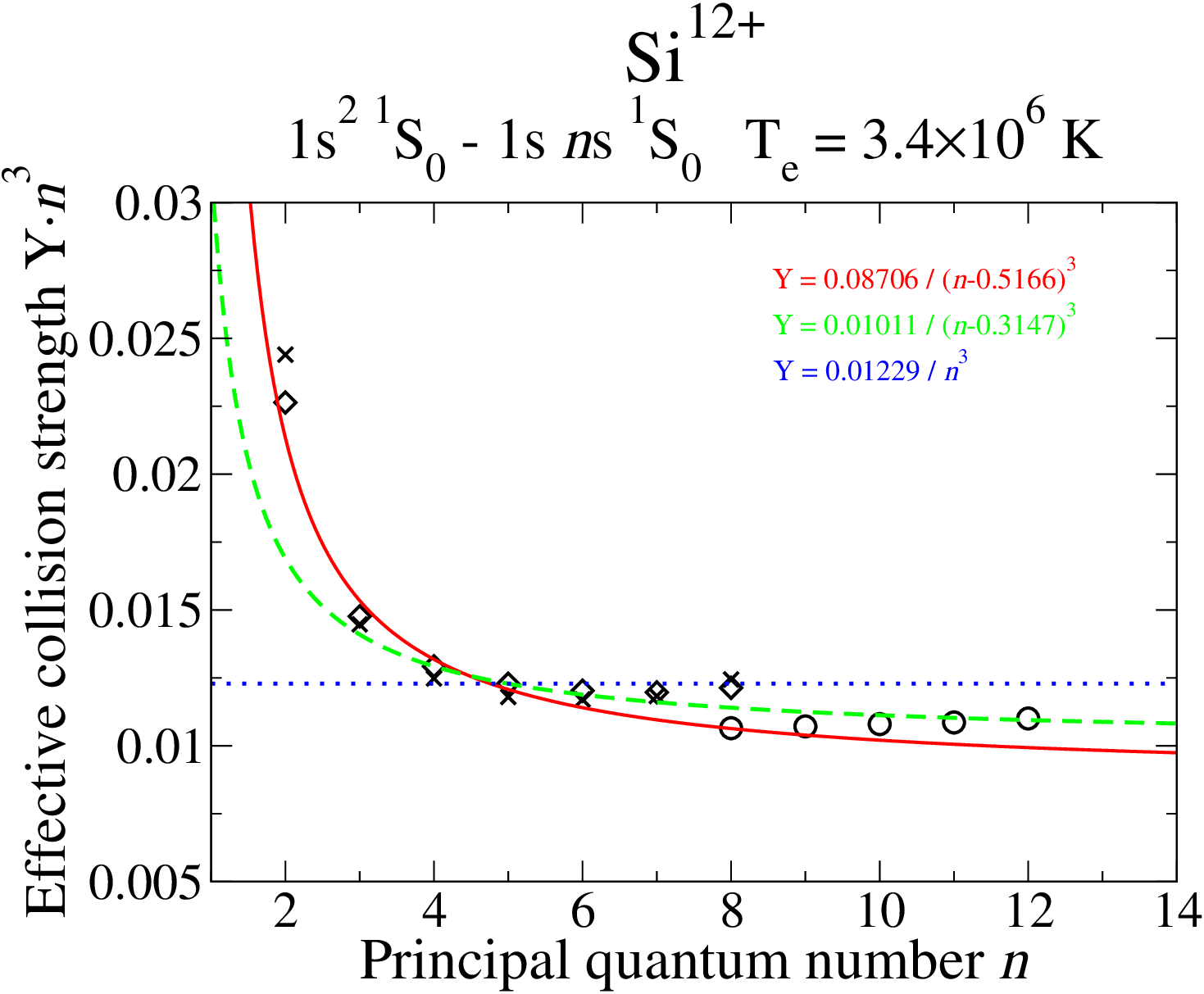}
}\,
\subfigure{
  \includegraphics[width=0.45\textwidth]{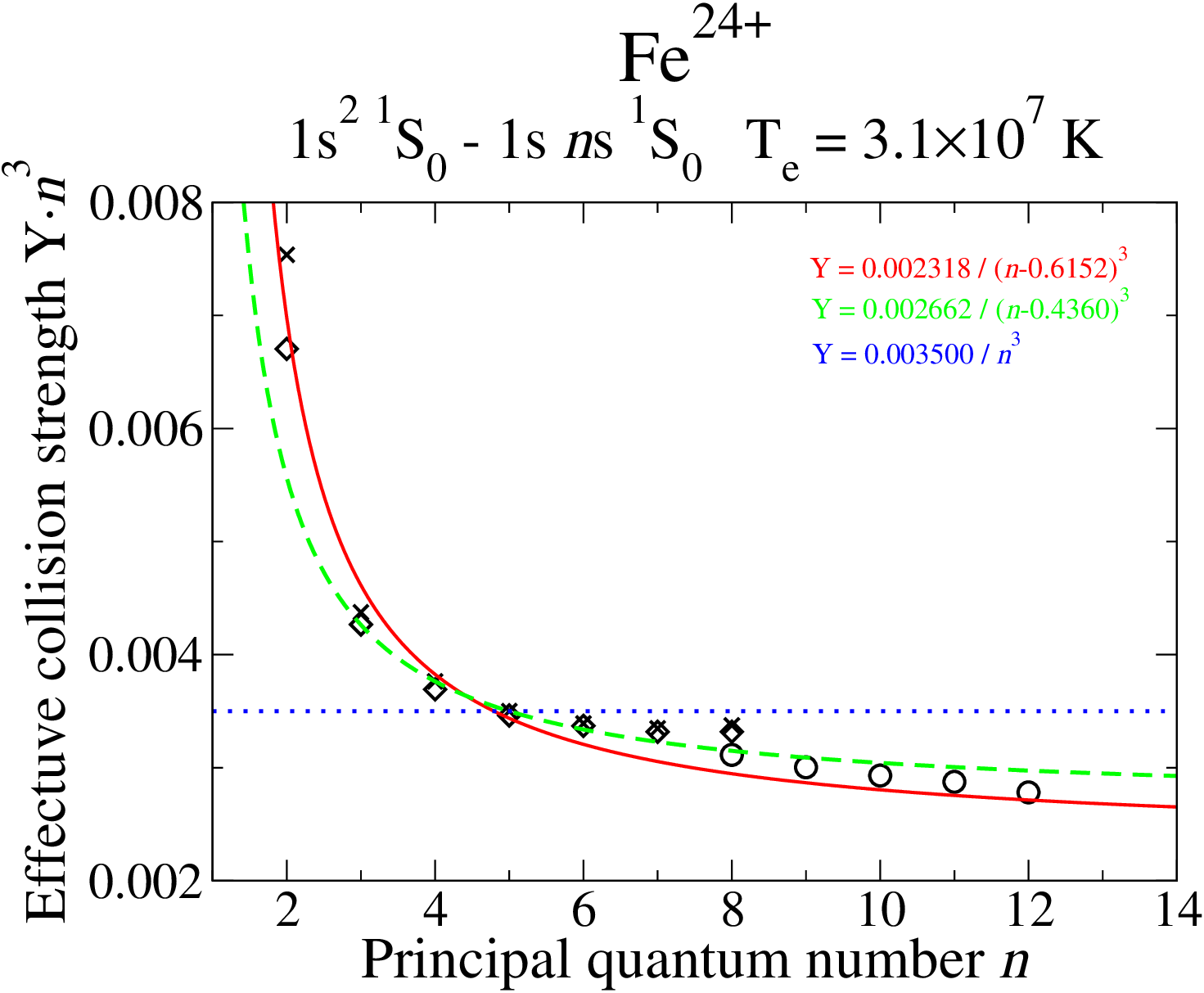}
}\\
\caption{Electron-impact excitation effective collision strengths  $(\times\, n^3)$ for 
  the Born transition 
  $\mathrm{1s^2\,^1S_0}-\mathrm{1s}n\mathrm{s\,^1S_0}$
  of the ions $\mathrm{Si}^{12+}$ and $\mathrm{Fe}^{24+}$ 
  versus the principal quantum number $n$ around peak abundance temperature.
  $\times$: $R$-matrix results;
  $\diamond$: distorted wave results with basis set $n=1-8$;
  $\circ$: distorted wave results with basis set $n=8-12$;
  solid line: least-squares fit using points $n=2-5$;
  dashed line: extrapolation using the last two points;
  dotted line: extrapolation using the last point; see text.}
\label{fig:upsvsn_ns1S}
\end{figure*}

Figure~\ref{fig:upsvsn_ns1S} shows the comparison between the $\Upsilon$
calculated with the $R$-matrix method, with the distorted wave method using both basis sets, 
and for the three extrapolation models, all at the peak abundance temperature.
For $n \ge 4$, the $R$-matrix and distorted wave results agree below~$10\%$.

\begin{figure}[!hbtp]
\centering
\includegraphics[width=0.45\textwidth]{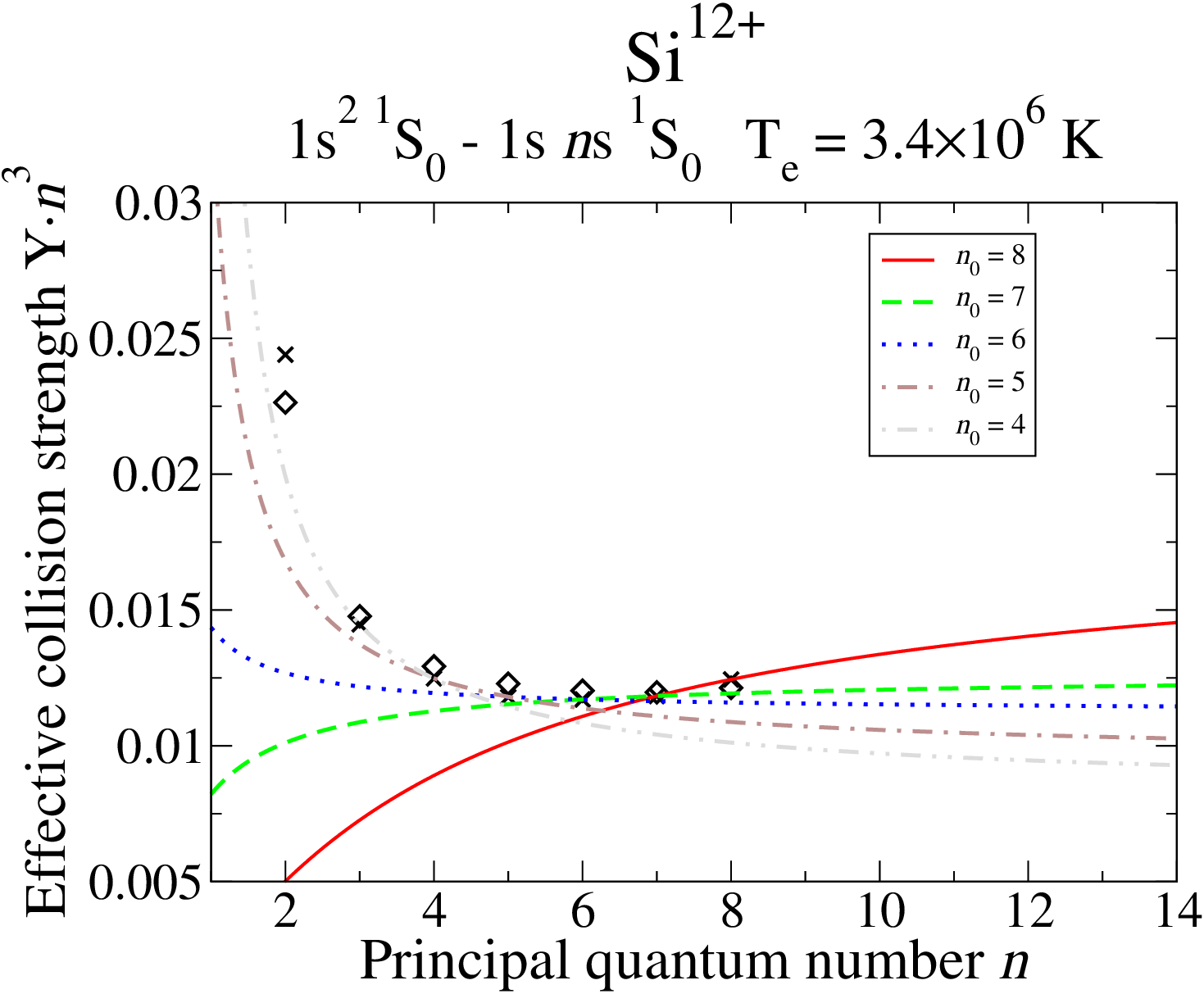}
\caption{Extrapolation curves for the $\Upsilon\times\, n^3$ displayed in 
  fig. \ref{fig:upsvsn_ns1S}, taking different extrapolation
  points $n_0$.
  $\times$: $R$-matrix results;
  $\diamond$: distorted wave results.}
\label{fig:upsvsnmulti_ns1S}
\end{figure}

To test the validity of the extrapolation rules for this type of transition
we show again in Table \ref{tab:params_ns1S} the calculated parameters for
the three methods with different reference points $n_0$ 
for $\mathrm{Si}^{12+}$.
Fig. \ref{fig:upsvsnmulti_ns1S} shows the analytic curves.

\begin{table}[hbt]
\caption{\label{tab:params_ns1S} Fitting parameters for the extrapolation
of the $\Upsilon$ at high-$n$ for the dipole electric transition 
of $\mathrm{Si}^{12+}$ $\mathrm{1s^2\,^1S_0}-\mathrm{1s}n\mathrm{s\,^1S_0}$
at a temperature of $T=3.4 \times 10^6 \kelvin$,
for different extrapolation reference points $n_0$.}
\begin{center}
\begin{footnotesize}
\begin{tabular}{|r|rr|rr|r|}
\hline
$n_0$ & \multicolumn{2}{c|}{Linear fit} &
\multicolumn{2}{c|}{Two point} & \multicolumn{1}{c|}{One point} \\
  & \multicolumn{1}{c}{$A$} & \multicolumn{1}{c|}{$\alpha$} &
    \multicolumn{1}{c}{$A$} & \multicolumn{1}{c|}{$\alpha$} &
    \multicolumn{1}{c|}{$A$} \\
\hline
  4 & $0.00721$ & $-0.6519$ & $0.00832$ & $-0.5050$ & $0.01248$  \\ 
  5 & $0.00794$ & $-0.5872$ & $0.00953$ & $-0.3445$ & $0.01180$  \\ 
  6 & $0.00865$ & $-0.5166$ & $0.01126$ & $-0.0778$ & $0.01171$  \\ 
  7 & $0.00934$ & $-0.4435$ & $0.01263$ & $ 0.1538$ & $0.01183$  \\ 
  8 & $0.01025$ & $-0.3421$ & $0.01813$ & $ 1.0691$ & $0.01244$  \\ 
\hline
\end{tabular}
\end{footnotesize}
\end{center}
\end{table}

For this transition the two-point model gives very similar results for 
extrapolation with $n_0=5,6,7$.
The two-point model with $n_0=8$ is slightly different, but the last point
of the calculation should not be considered a good reference for the 
extrapolation.
The error of the two-point model extrapolation from $n_0=5,6,7$ with respect
to the calculated $R$-matrix data is less than $10\%$.
For Born-allowed transitions, we recommend a two-point extrapolation model as the
most accurate.
The $\alpha$ parameters calculated with the two-point model are also close 
to zero for $n_0=6,7$, and so here the one-point extrapolation also gives accurate results.

In general, the calculations with the oversimplified atomic structure give worse results
than the extrapolated ones.
We do not recommend such simplifications at all for estimation purposes.

\subsection{Forbidden transitions}
\label{subsec:forbidden}

These transitions are characterised by zero limit points, electric dipole and
plane-wave Born \citep{burgess1992}.
They can arise for spin-changing and/or multiple-electron jump transitions.
They are infrequent because of configuration and spin-orbit
mixing.
For high impact energies, the collision strength decays with a power law
$\Omega \sim E^{-\gamma}$, with $\gamma$ close to two.
This rapid decay makes the resonance enhancement large, particularly at
low temperatures.

\begin{figure*}[hbtp]
\centering
\subfigure{
  \includegraphics[width=0.45\textwidth]{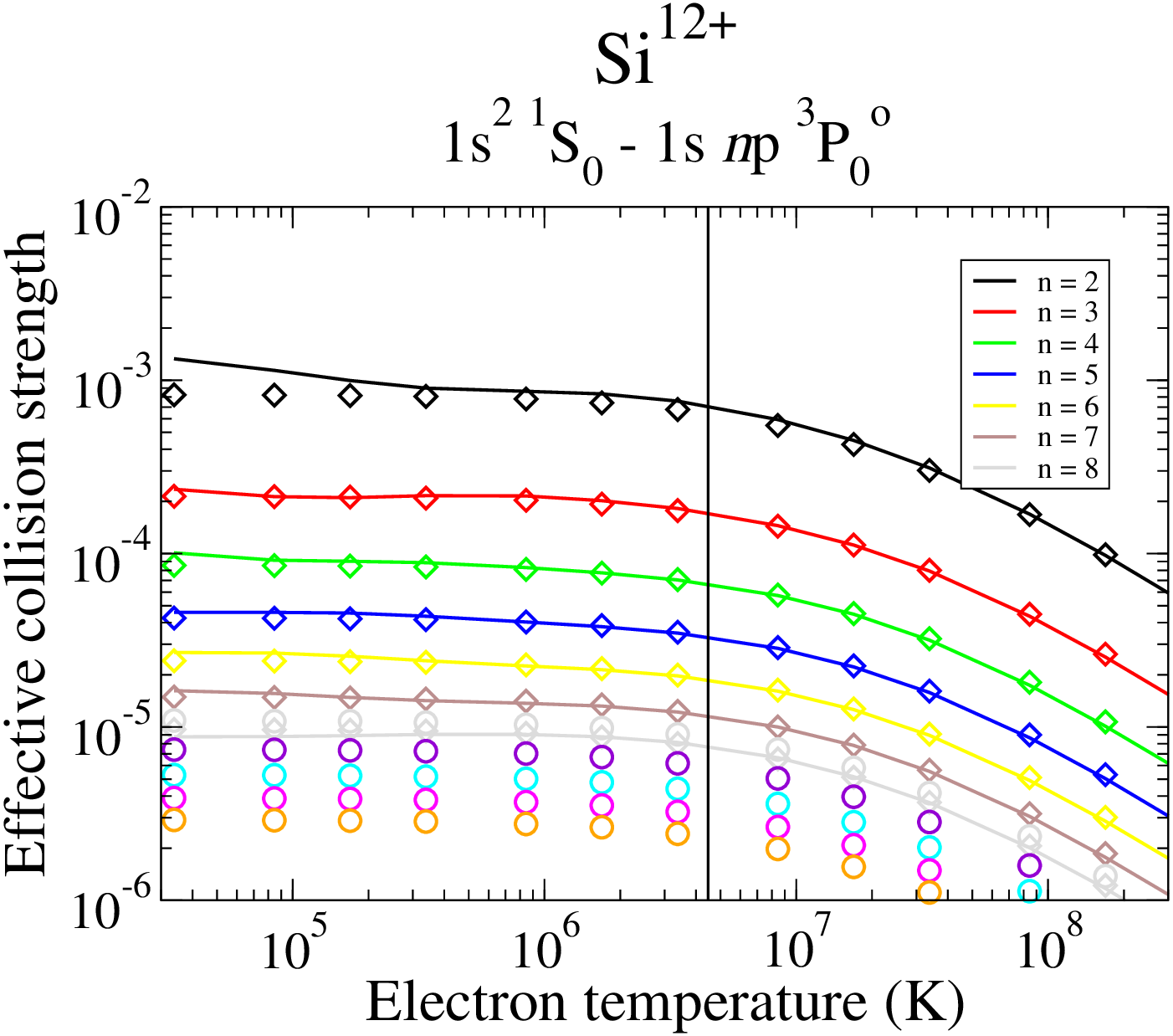}
}\,
\subfigure{
  \includegraphics[width=0.45\textwidth]{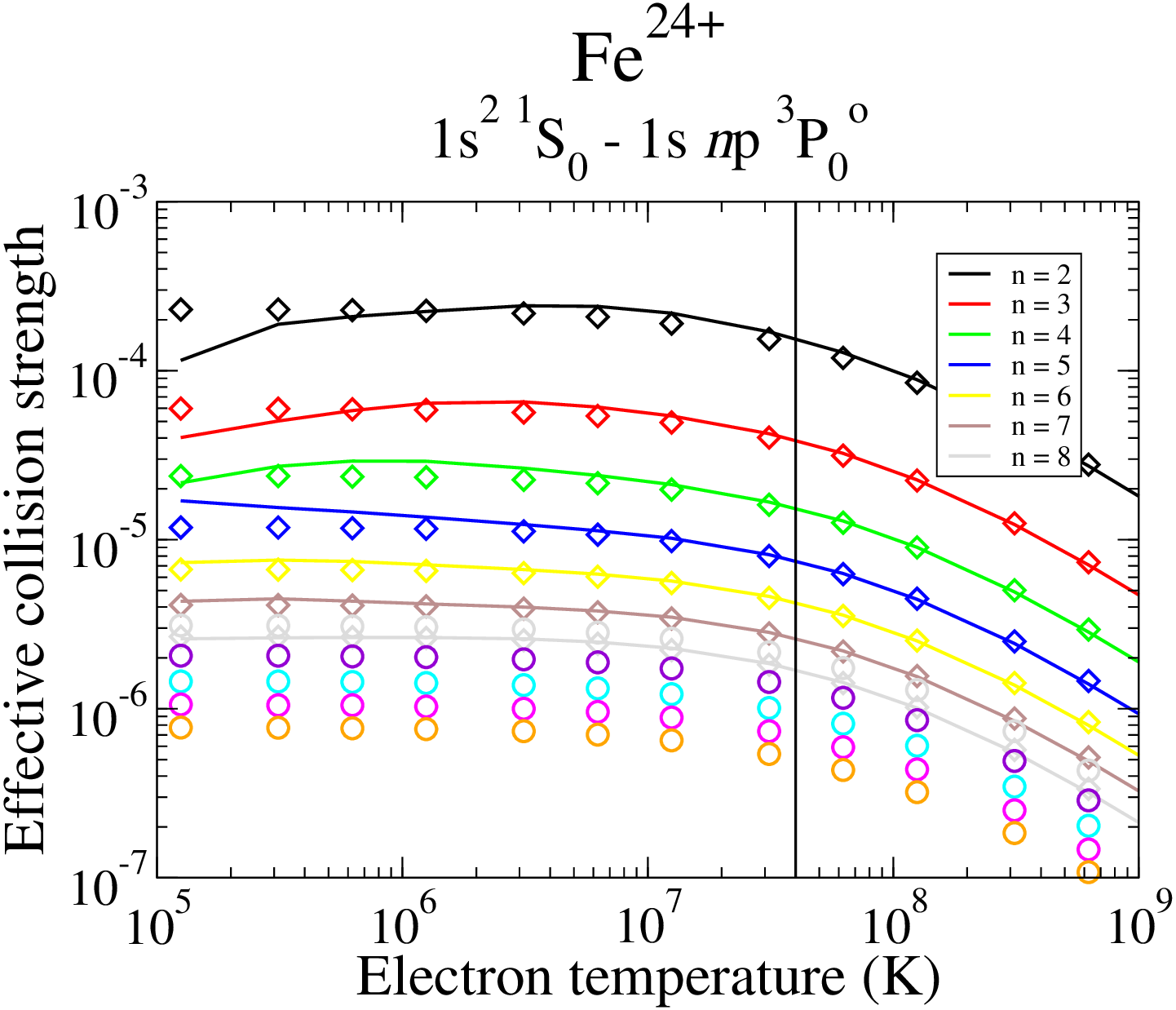}
}\\
\caption{Electron-impact excitation effective collision strengths for 
  the forbidden transition 
  $\mathrm{1s^2\,^1S_0}-\mathrm{1s}n\mathrm{p\,^3P_0}$
  of the ions $\mathrm{Si}^{12+}$ and $\mathrm{Fe}^{24+}$.
  Curved lines: $R$-matrix; 
  $\diamond$: distorted wave calculation basis $n=1-8$;
  $\circ$: distorted wave calculation basis $n=8-12$;
  vertical line: peak abundance temperature for electron collisional 
  plasmas~\citep{mazzotta1998}.}
\label{fig:ups_np3P0}
\end{figure*}

Figure~\ref{fig:ups_np3P0} shows the effective collision strengths for the
pure spin-change transition 
$\mathrm{1s^2\,^1S_0}-\mathrm{1s}n\mathrm{p\,^3P_0}$ for the ions
$\mathrm{Si}^{12+}$ and $\mathrm{Fe}^{24+}$.
As expected, at low temperatures there are differences up to a factor of two
between distorted wave and $R$-matrix due to the resonances.

\begin{figure*}[hbtp]
\centering
\subfigure{
  \includegraphics[width=0.45\textwidth]{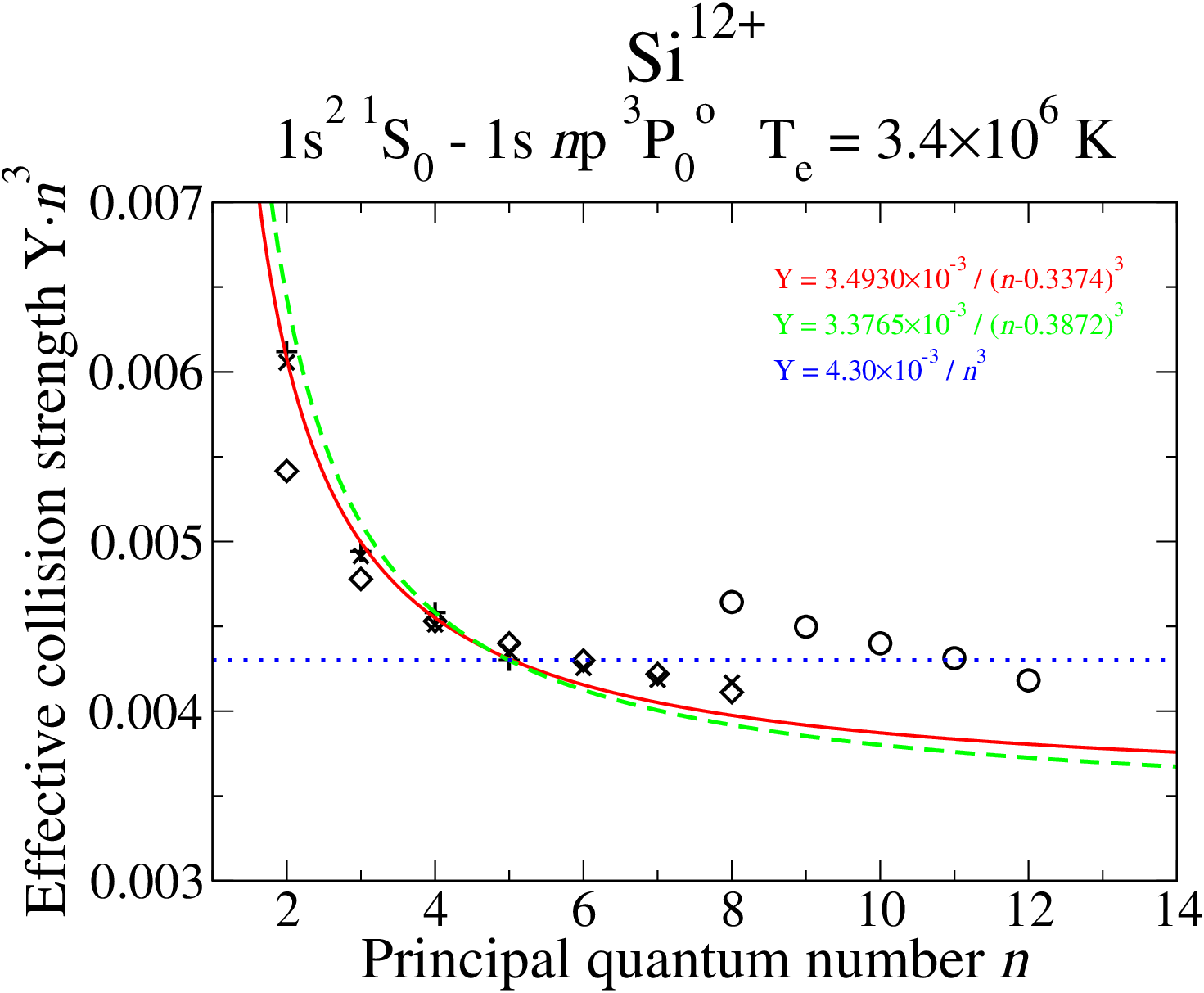}
}\,
\subfigure{
  \includegraphics[width=0.45\textwidth]{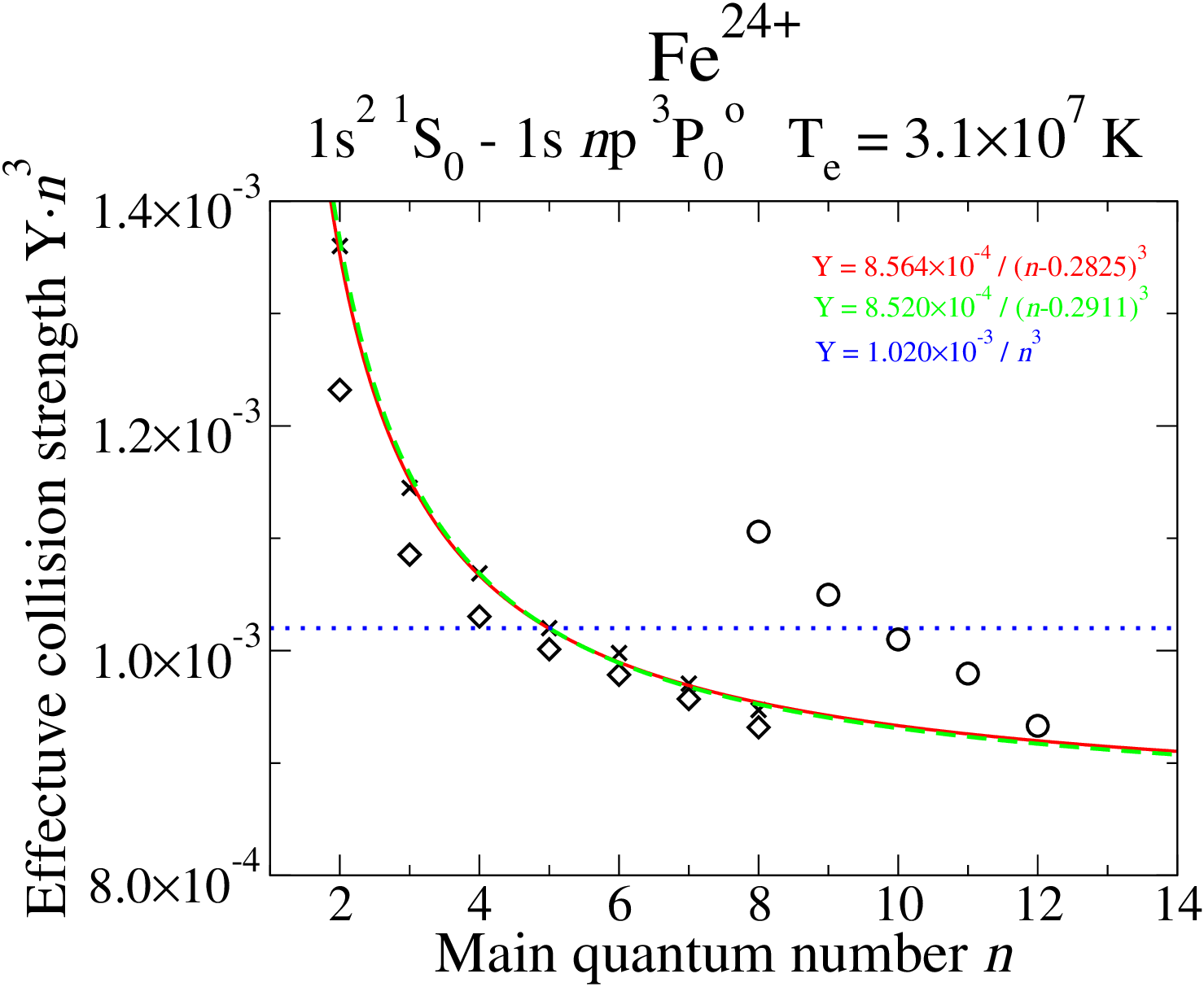}
}\\
\caption{Electron-impact excitation effective collision strengths $(\times\, n^3)$ for 
  the forbidden transition 
  $\mathrm{1s^2\,^1S_0}-\mathrm{1s}n\mathrm{p\,^3P_0}$
  of the ions $\mathrm{Si}^{12+}$ and $\mathrm{Fe}^{24+}$ 
  versus the principal quantum number $n$ around peak abundance temperature.
  $\times$: $R$-matrix results;
  $\diamond$: distorted wave results with basis set $n=1-8$;
  $\circ$: distorted wave results with basis set $n=8-12$;
  solid line: least-squares fit using points $n=2-5$;
  dashed line: extrapolation using the last two points;
  dotted line: extrapolation using the last point; see text.}
\label{fig:upsvsn_np3P0}
\end{figure*}

Figure~\ref{fig:upsvsn_np3P0} shows the comparison between the $\Upsilon$
calculated with the $R$-matrix method, the distorted wave methods with both basis sets, 
and the three extrapolation models, at the peak abundance temperature.
In this case, the  models fit quite well the data for $n=6-8$.
For these transitions, the value of the parameter $\alpha$ is quite large,
and the results of models type 2 and 3 differ significantly.
The calculations with the simplified atomic structure again poorly reproduce
the data. 

We show again in Table \ref{tab:params_np3P0} the different extrapolation
parameters calculated with the three methods and different reference 
point $n_0$.
Figure \ref{fig:upsvsn_np3P0} shows the extrapolation curves.
Calculated values for the parameters $A$ and $\alpha$ are quite similar for 
the different reference points $n_0=5,6,7$ if the same model is used,
one or two-point.
On the other hand, if we compare the results obtained with same $n_0$ but
different models, they are quite different.
The one-point model does not correctly reproduce the behaviour  of
the $R$-matrix results for high-$n$.
For this type of transition we recommend a two-point model.

\begin{table}[hbt]
\caption{\label{tab:params_np3P0} Fitting parameters for the extrapolation
of the $\Upsilon$ at high-$n$ for the dipole electric transition 
of $\mathrm{Si}^{12+}$ $\mathrm{1s^2\,^1S_0}-\mathrm{1s}n\mathrm{p\,^3P_0}$
at a temperature of $T=3.4 \times 10^6 \kelvin$,
for different extrapolation reference points $n_0$.}
\begin{center}
\begin{footnotesize}
\begin{tabular}{|@{}r|rr@{}|rr@{}|@{}r@{}|}
\hline
$n_0$ & \multicolumn{2}{c|}{Linear fit} &
\multicolumn{2}{c|}{Two point} & \multicolumn{1}{c|}{One point} \\
  & \multicolumn{1}{c}{$A$} & \multicolumn{1}{c|}{$\alpha$} &
    \multicolumn{1}{c}{$A$} & \multicolumn{1}{c|}{$\alpha$} &
    \multicolumn{1}{c|}{$A$} \\
\hline
  4 & $3.451 \cdot 10^{-3}$ & $-0.3105$ & 
      $3.541 \cdot 10^{-3}$ & $-0.3105$ & $4.512 \cdot 10^{-3}$  \\
  5 & $3.553 \cdot 10^{-3}$ & $-0.3163$ & 
      $3.774 \cdot 10^{-3}$ & $-0.2311$ & $4.350 \cdot 10^{-3}$  \\
  6 & $3.626 \cdot 10^{-3}$ & $-0.2981$ & 
      $3.820 \cdot 10^{-3}$ & $-0.2119$ & $4.255 \cdot 10^{-3}$  \\
  7 & $3.670 \cdot 10^{-3}$ & $-0.2858$ & 
      $3.792 \cdot 10^{-3}$ & $-0.2260$ & $4.185 \cdot 10^{-3}$  \\
  8 & $3.725 \cdot 10^{-3}$ & $-0.2690$ & 
      $4.052 \cdot 10^{-3}$ & $-0.0748$ & $4.168 \cdot 10^{-3}$  \\
\hline
\end{tabular}
\end{footnotesize}
\end{center}
\end{table}

\begin{figure}[hbtp]
\centering
\includegraphics[width=0.45\textwidth]{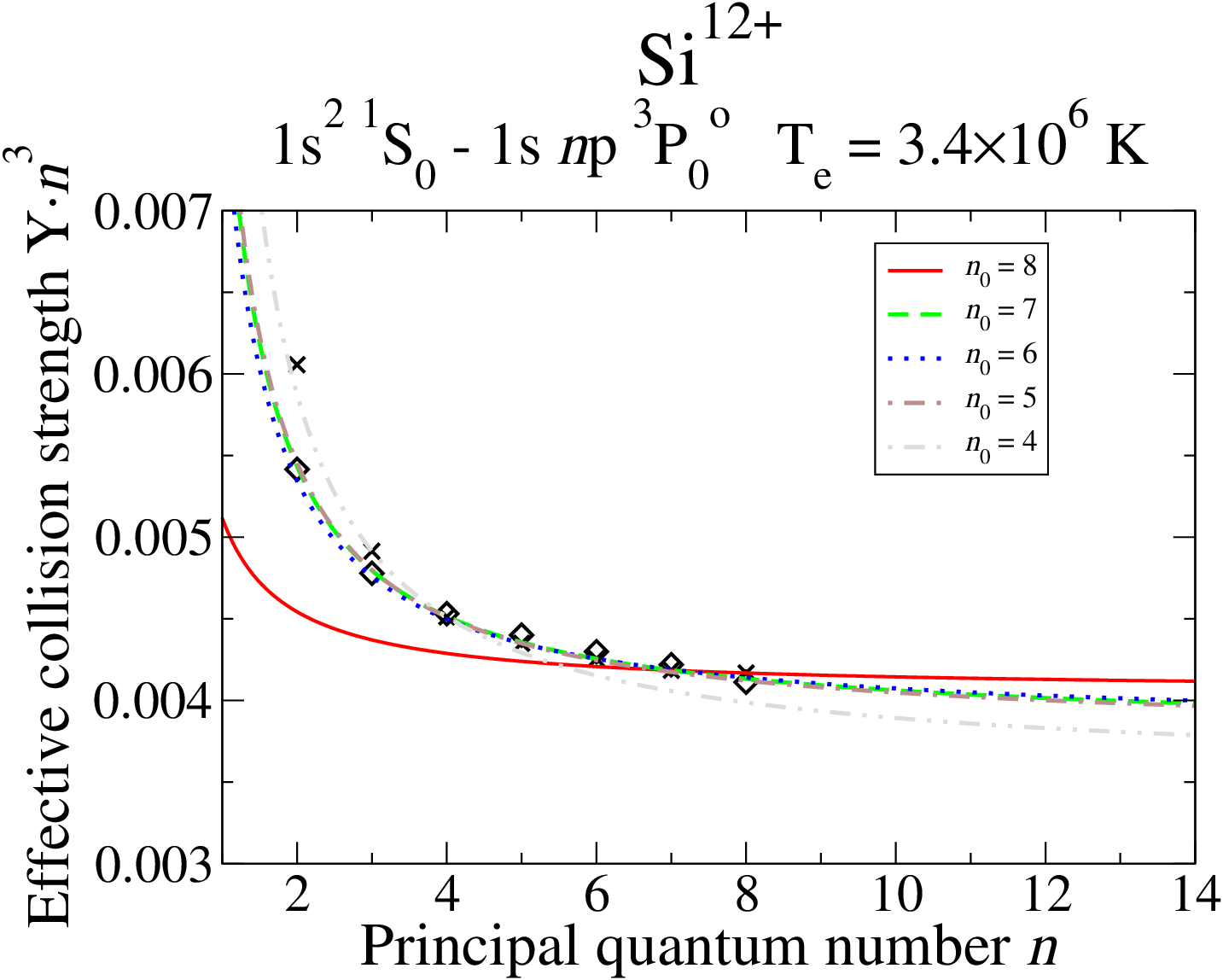}
\caption{Extrapolation curves for the $\Upsilon$ displayed in Fig. \ref{fig:upsvsn_np3P0}, taking different extrapolation
  points $n_0$.
  $\times$: $R$-matrix results.}
\label{fig:upsvsnmulti_np3P0}
\end{figure}

\subsection{Low-charged ions}
\label{subsec:lowcharge}

The $n^{-3}$  behaviour of the collision strengths is fulfilled if we are in 
the limit of Rydberg atom, that is when the interaction of the core with the
active electron can be considered a one-body Coulomb one.
For lower-charged atoms the electron-electron interaction is of the same order 
as the nucleus-electron one.
For such ions this Rydberg atom limit is reached at a higher value
of the principal quantum number $n$.
In principle, the above extrapolation rules should work in a lower-charged ion, 
but the reference point $n_0$ should be high enough for it to be considered a Rydberg 
atom.
This means that reference data for extrapolation are required up 
to $n \approx 8-10$. 
This is a rather large $R$-matrix calculation, because of the large box-size.
In addition, the coupling with the continuum increases as the charge
decreases.
So a good quality calculation for high-$n$ for a low-charged ion must
include pseudostates in the CI/CC expansions.

\begin{figure*}[hbtp]
\centering
\subfigure{
  \includegraphics[width=0.45\textwidth]{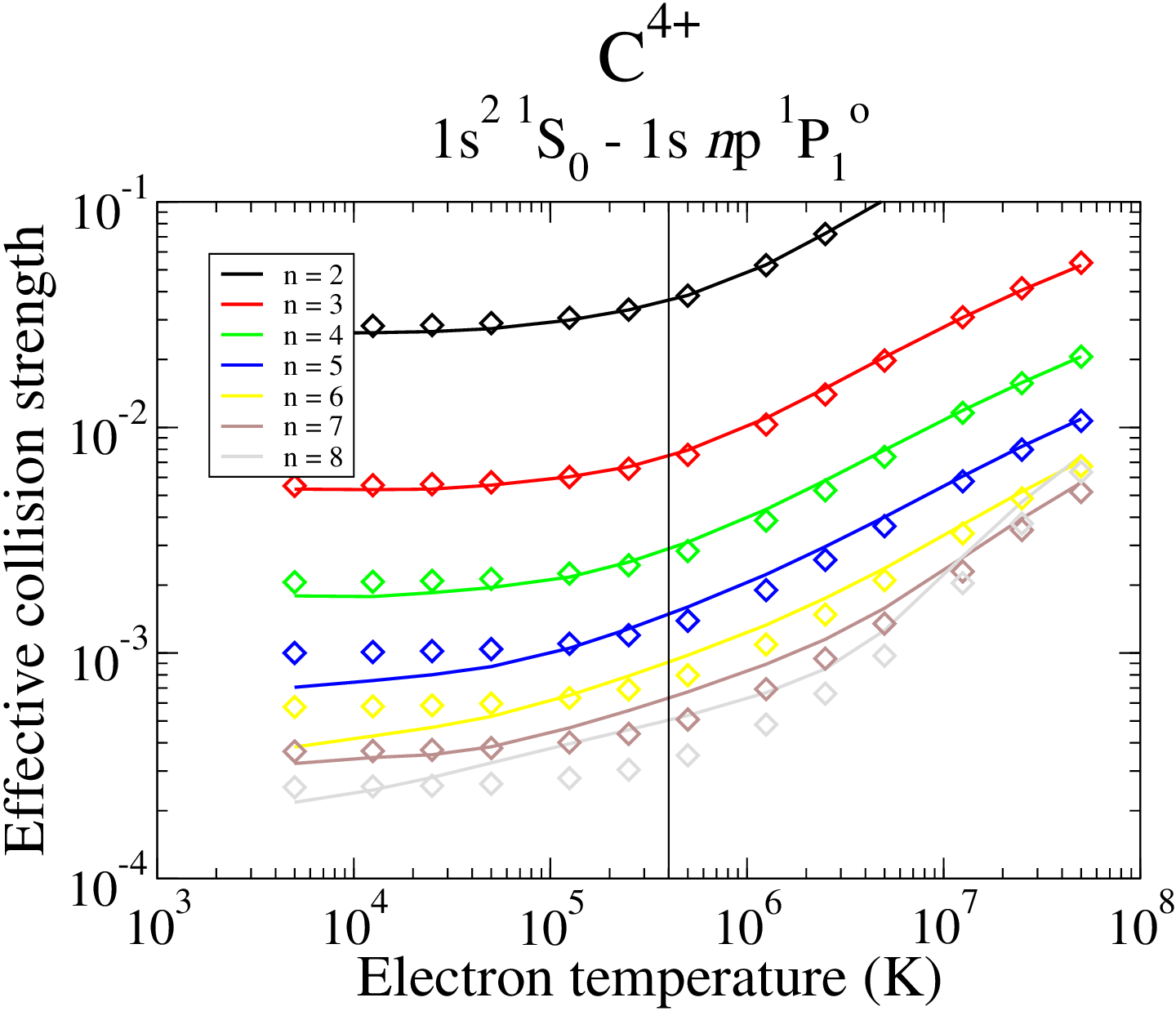}
}\,
\subfigure{
  \includegraphics[width=0.45\textwidth]{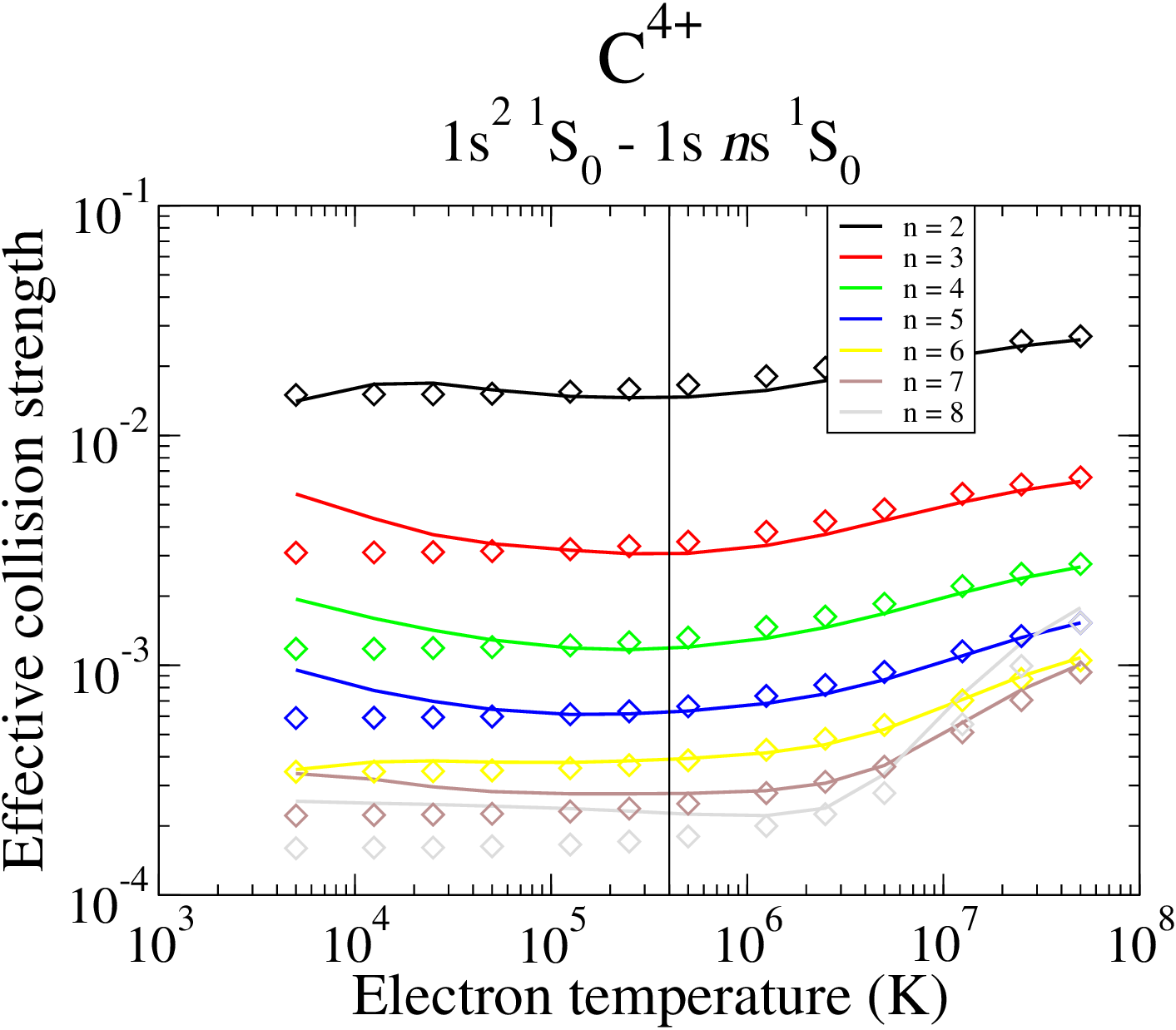}
}\\
\caption{Electron-impact excitation effective collision strengths for 
  dipole allowed and Born transitions 
  $\mathrm{1s^2\,^1S_0}-\mathrm{1s}n\mathrm{p\,^1P_1^{o}}$
  $\mathrm{1s^2\,^1S_0}-\mathrm{1s}n\mathrm{s\,^1S_0}$
  of the ion $\mathrm{C}^{4+}$.
  Curved lines: $R$-matrix; 
  $\diamond$: distorted wave calculation basis $n=1-8$;
  vertical line: peak abundance temperature for electron collisional 
  plasmas~\citep{mazzotta1998}.}
\label{fig:ups_c4}
\end{figure*}

Figure~\ref{fig:ups_c4} shows the effective collision strengths for dipole and 
Born allowed transitions of  $\mathrm{C}^{4+}$.
The background collision strength falls off as $z^{-2}$ while, initially,
the resonance strength is independent of $z$, although at sufficiently high charge
radiation damping usually starts to reduce the resonance contribution.
So in $\mathrm{C}^{4+}$ the resonance enhancement of the effective
collision strengths at low temperatures is seen to be relatively smaller than for 
$\mathrm{Si}^{12+}$ and $\mathrm{Fe}^{24+}$.

The effective collision strengths for the last $n$-shells included in the
basis set show irregular behaviour.
This is caused by the loss of quality for the description of the highest
excited states.
In low-charged ions the inaccuracy in the description of the atomic structure
is larger due to stronger coupling with more highly-excited bound states and the continuum,
which we neglect.
For the present test calculations we did not include pseudostates in the 
description of $\mathrm{C}^{4+}$ atomic structure or the close-coupling 
expansion. Thus, we do not recommend using this data in preference to 
$R$-matrix with pseudo-states data, rather it is a guide to extrapolating
data in low-charge ions.
The uncertainty associated with an inaccurate atomic structure
due to, for example, the lack of pseudostates in the case of a low-charged ion,
generates a much larger error than that associated with the use of an extrapolation formula.

\subsection{Other sequences}
\label{sec:other}

As explained above, 
the behaviour of the collision strengths with respect to the quantum number 
tends to the form given by (\ref{eq:extrap}) when the active electron is highly enough
excited, so the ion can be considered a Rydberg one.
We need to consider transitions between Rydberg states with a difference in 
$n$-values between the active electron and the highest core-electron of at least two.
For low-charged ions, this difference in $n$ may be necessary to be increased to
up to four.
In $\mathrm{H}$- and $\mathrm{He}$-like sequences 
the  $\sim n^{-3}$ behaviour applies starting from $n=5$, as shown before.
For the $\mathrm{Li}$- and $\mathrm{Be}$-like sequences the starting shell 
is  $n=6$ and for $\mathrm{Na}$- and $\mathrm{K}$-like the $n=7$.

As the number of electrons increases, the size of the basis set required to 
obtain  accurate results for the  excited shells increases significantly.
The complication of the core calculations increases and the application of
the extrapolation rules becomes impossible.
The present extrapolations  provide good estimates only for the
$\mathrm{H}$- and $\mathrm{He}$-like sequences.
For other sequences they may not be valid, 
and cannot be easily tested.

\section{Comparisons with observations}
\label{sec:comparison}

\begin{table}[!htbp]
\begin{center}
\caption{Experimental and theoretical ratios (ergs) for \ion{Si}{xiii}.
The observed values are from \citet{kepa2006}; K06.   
We list our theoretical values, together with those reported by K06.}
\label{tab:ratiossi12}
\footnotesize
\begin{tabular}{lllllll}
\hline\hline\noalign{\smallskip}
  & Observed      & 5 MK  & 10 MK & 15 MK & 25 MK &     \\
\noalign{\smallskip}\hline\noalign{\smallskip}

4p/3p & 0.28--0.448  & 0.29  & 0.33  & 0.345 & 0.356 &     \\
   &              & 0.272 & 0.308 & 0.322 & 0.333 & K06 \\
5p/3p & 0.131--0.205 & 0.129 & 0.154 & 0.164 & 0.170 &     \\
   &              & 0.115 & 0.138 & 0.147 & 0.154 & K06 \\
6p/3p & 0.071--0.089 & 0.066 & 0.084 & 0.090 & 0.094 &     \\
   &              & 0.060 & 0.074 & 0.080 & 0.085 & K06 \\
7p/3p &      -       & 0.044 & 0.051 & 0.055 & 0.058 &     \\
8p/3p &      -       & 0.026 & 0.033 & 0.036 & 0.0375 &    \\
9p/3p &      -       & 0.0175 & 0.0225 & 0.0245 & 0.0255 & \\
10p/3p &     -       & 0.012 & 0.0154 & 0.0168 & 0.0175 &  \\

\noalign{\smallskip}\hline
\end{tabular}
\normalsize
\end{center}
\end{table}

\begin{table}[!htbp]
\begin{center}
\caption{Experimental and theoretical ratios (ergs) for \ion{S}{xv}.
The observed values are from \citet{kepa2006}; K06.
We list our theoretical values, together with those reported by K06.}
\label{tab:ratioss14}
\footnotesize
\begin{tabular}{lllllll}
\hline\hline\noalign{\smallskip}
  & Observed     & 5 MK & 10 MK  & 15 MK & 25 MK & \\
\noalign{\smallskip}\hline\noalign{\smallskip}

2p/3p & 10.5--11.57  & 12.45 &  7.73 & 6.6   & 5.85  &     \\
   &              & 16.3  &  10.6 & 9.3   & 8.55  & K06 \\
4p/3p & 0.27--0.912  & 0.27  & 0.318 & 0.337 & 0.354 &     \\
   &              & 0.25  & 0.296 & 0.314 & 0.328 & K06 \\
5p/3p &  -           & 0.11  & 0.144 & 0.157 & 0.168 &     \\
   &              & 0.10  & 0.13  & 0.141 & 0.151 & K06 \\
6p/3p &  -           & 0.057 & 0.077 & 0.085 & 0.092 &     \\
   &              & 0.052 & 0.069 & 0.076 & 0.083 & K06 \\
7p/3p & 0.045--0.417 & 0.034 & 0.046 & 0.051 & 0.056 &     \\
   &              & 0.030 & 0.042 & 0.046 & 0.051 & K06 \\
8p/3p & 0.028--0.167 & 0.021 & 0.0297 & 0.033 & 0.036 &    \\
   &              & 0.019 & 0.027 & 0.030 & 0.033 & K06 \\
9p/3p & 0.020--0.142 & 0.014 & 0.021 & 0.022 & 0.024 &     \\
   &              & 0.013 & 0.019 & 0.021 & 0.023 & K06 \\
10p/3p &   -         & 0.010 & 0.014 & 0.015 & 0.016 &     \\

\noalign{\smallskip}\hline
\end{tabular}
\normalsize
\end{center}
\end{table}

In Tables \ref{tab:ratiossi12} and \ref{tab:ratioss14} we show a comparison of 
the line ratios calculated with the extrapolation rules of the present work 
with the observed ones of \citet{kepa2006} for the soft X-rays detected by 
RESIK coming from highly-excited states of \ion{Si}{XIII} and \ion{S}{XV} in 
solar flares.
The transitions involved in the line ratios are dipole allowed.

We have used as our starting point for the extrapolation calculations
the $n=5$ $R$-matrix data of \citet{whiteford2001}.
We have used a two-point extrapolation model and
a reference point $n_0=5$, and we have obtained extrapolated
$\Upsilon$ from the ground state up to $n=10$.
In the online material we provide in CDS format the extrapolated values
of energies, radiative parameters $gf$, and electron-impact excitation 
effective collision strengths $\Upsilon$ obtained with the extrapolation 
rules described here.
The modelling has been carried out by converting these data into CHIANTI
\citet{delzanna2015c} format and using the CHIANTI population solver to obtain 
the level populations.

Our theoretical line ratios are quite close (within $10\%$ for most cases) 
with those estimated by \citet{kepa2006}. 
Regarding the observed values, \citet{kepa2006} report a range. 
The lower values correspond to the peak and gradual phase of the flares. 
They are within the theoretical values in the $5-10\,\mathrm{MK}$ range. 
The higher values, instead, correspond to the early impulsive phase, and are 
in most cases outside the range of the theoretical values.

\section{Discussion and conclusions}
\label{sec:conclusions}

We carried out several new calculations, both $R$-matix and distorted wave, for the
electron-impact excitation  of $\mathrm{H}$- and $\mathrm{He}$-like ions.
We have shown the dependence of the effective collision strengths $\Upsilon$
with respect to the principal quantum number $n$ for several transition types of 
some benchmark ions of the $\mathrm{He}$-like isoelectronic sequence.
We tested three models to reproduce the behaviour of the $\Upsilon(n)$
and extrapolate them to more highly-excited states.

In general, the extrapolation rules do not give such accurate values for the
effective collision strengths as explicit calculations do; instead,
they give an approximation that can be used for estimation purposes in modelling.
Clearly, it is first necessary to have a good starting calculation before 
applying the extrapolations to the data.
We note that $R$-matrix results become increasingly uncertain for the highest energy states 
included in the CI and CC expansions of the target, due to their lack of convergence \citep{fernandez-menchero2015b}.
The description of the atomic structure, energy levels and radiative data, and
the corresponding effective collision strengths, increasingly lose accuracy as we
approach  the last states included in the basis expansions.
The same happens with the distorted wave calculations.
Even when the distorted wave results are not affected by the coupling
or resonances, which are included in the $R$-matrix ones, 
the description of the atomic structure is its main limitation.
In consequence, such calculations may not give more accurate results for high-$n$ than
the extrapolation rules combined with accurate calculations for lower, but
sufficiently excited, states.

Among the three extrapolation models considered, we recommend the second one.
The Model 2 two-point extrapolation provides results that are closer to the $R$-matrix
data, and reproduces the behaviour at high-$n$.
Nevertheless it is not a completely general rule, and each particular 
transition type should be analysed to check which is the most accurate 
extrapolation method for the ion.
We do not recommend Model 1 the least-squares fitting for several reasons.
First, having just four points is not enough for a good-quality fitting.
Second, the first data points $n=2,3$ have not yet reached the required
$n^{-3}$ behaviour. This occurs because they can not be considered Rydberg 
levels: they interact more strongly with the core and  resonances play a 
larger role for excitation to these shells.
Finally, the fitting method requires a considerably larger computational
effort to obtain a result that may be worse than the simpler methods.

The Model 3 one-point extrapolation gives, in general, a poorer estimate than
the two-point one, and the computational work is not substantially reduced.
On the other hand, sometimes the parameters estimated with the one-point rule
are more stable with respect to the reference point $n_0$ than the two-point one,
for example, the dipole transitions shown here.

Due to the inherent uncertainty in data for the most highly-excited states,
we do not recommend  extrapolating  the $\Upsilon$ from the
last $n$-shell, but suggest dropping form the extrapolation the last one or two
$n$-values.
Also, the  $n^{-3}$ behaviour applies  only from a certain excited shell.
The extrapolation has to start from a level when the atom can be considered
as a Rydberg one.
That is  at least approximately two shells higher than the last occupied one
($n=3$ for $\mathrm{H}$- and $\mathrm{He}$-like sequences).

We also do not recommend the use of models with an oversimplified atomic structure
so as to reach high-$n$ shells.
The extrapolation from an accurate calculation to lower-$n$ provides in general
a better estimate than a larger explicit calculation with a poorer atomic structure.

As an example application, we have compared lines ratios obtained with the presented
extrapolation-rule Model 2
with observations of X-rays by RESIK.
We obtain good agreement with the observed \ion{Si}{xiii} and \ion{S}{xv} 
ratios during the peak phase of solar flares, but the values during the 
impulsive phase are still outside the theoretical range.  
It is expected that during the impulsive phase non-equilibrium effects are 
present. 
We investigated whether a non-Maxwellian distribution such as a  
$\kappa$-distribution was able to increase the ratios, but did not find significant 
increases. 
We are currently investigating other possible causes.

For photoionised plasmas the ions  exist at temperatures lower than the
peak abundance in an electron collision dominated plasma.
At these lower temperatures, the $\Upsilon$ are affected more by resonance enhancement,
and perhaps radiation damping thereof,
and can depend greatly on the position of these resonances, which are 
determined by the atomic structure.

Resonance effects can cause the 
$\Upsilon$ to deviate from  the $n^{-3}$ rule.
The extrapolation rules should therefore only be applied starting from a 
higher excited shell, so these effects are minimised.
This occurs  even if the calculations are perfectly accurate at low temperatures.
In these case, we suggest to start the extrapolation at least from four shells 
above the last occupied ($n=5$ for $\mathrm{H}$- and $\mathrm{He}$-like 
sequences).

We estimate that the accuracy of the extrapolation Model 2 is 
approximately  $20\%$ for all 
transition types of moderately- and highly-charged ions, and approximately 
$50\%$ for low-charged ions.
This estimate  considers  the core calculations to be perfect and the extrapolation
carried out from a shell where the ion can be considered as Rydberg.
The inaccuracies in the core calculations could lead to larger errors
in some cases, especially for low-charged ions and/or low temperatures.

\begin{acknowledgements}
This work was funded by STFC (UK) through the 
University of Strathclyde UK APAP network grant ST/J000892/1
and the University of Cambridge  DAMTP astrophysics consolidated grant.
Luis Fern\'andez-Menchero thanks the National Science Foundation (USA)
for the grant PHY-1520970.
\end{acknowledgements}

\def\baselinestretch{1.0}
\bibliographystyle{aa}
\bibliography{references}


\end{document}